\documentclass[]{aa}

\usepackage{amssymb}
\usepackage{epsfig} 
\usepackage{graphicx} 
\usepackage{color} 
\usepackage{times} 
\usepackage{amsmath} 
\usepackage{amssymb} 
\usepackage{xspace} 
\usepackage[varg]{txfonts} 
\usepackage{enumitem} 
\usepackage{natbib} 
\usepackage{algorithm} 
\usepackage{algorithmic} 
\usepackage{subfloat} 
\usepackage{subfig} 
\usepackage[normalem]{ulem}
\usepackage{booktabs}
\usepackage{lscape}

\usepackage{lineno}

\def\d3k{{\displaystyle {\rm d}{\bf k} \over \displaystyle (2\pi)^3}}

\def\apjl{Apj Letters} 

\newcommand{\cgal}{\texttt{CGAL}\ } 
\newcommand{\Excursion}  {\mm{{\mathbb E}}}
\newcommand{\Mask}       {\mm{{\mathbb M}}}
\newcommand{\Rspace}     {\mm{{\mathbb R}}} 
\newcommand{\Sspace}     {\mm{{\mathbb S}}} 
 
\newcommand{\Homology}[1]{\mm{{\sf H}_{#1}}}
\newcommand{\Rank}[1]    {\mm{{\rm rank\,}{#1}}}
\newcommand{\Betti}[1]   {\mm{{\beta}_{#1}}}
\newcommand{\relBetti}[1]{\mm{{b}_{#1}}}
\newcommand{\Euler}      {\mm{\sf EC}} 
\newcommand{\relEuler}      {\mm{\sf EC_{\rm rel}}} 
 
\newcommand{\dime}[1]    {\mm{\rm dim\,}{#1}}
\newcommand{\ssx}        {\mm{\sigma}}
\newcommand{\tsx}        {\mm{\tau}}

\newcommand{\R}{\mathbb{R}}

\renewcommand{\S}{\mathbf{S}}
\newcommand{\T}{{\tt T}}

\newcommand{\x}{\mathbf{x}}

\newcommand{\y}{\mathbf{y}}

\newcommand {\mm}[1] {\ifmmode{#1}\else{\mbox{\(#1\)}}\fi}

\newcommand{\pivot}[1]       {\mm{\rm pivot}{({#1})}}

\newcommand{\Res}            {\mm{N}}

\newcommand{\Skip}[1]        {}

\begin{document}

\title{{Unexpected Topology of the Temperature Fluctuations in the Cosmic Microwave Background}}

\author{Pratyush Pranav\inst{1,2}
  \thanks{\emph{Present address:} Univ Lyon, ENS de Lyon, Univ Lyon1, CNRS, Centre de Recherche Astrophysique de Lyon UMR5574, F--69007, Lyon, France
    }
  \and Robert J. Adler\inst{2}
  \and Thomas Buchert\inst{1}
  \and Herbert Edelsbrunner\inst{3}
  \and Bernard J.T. Jones\inst{5}
  \and Armin Schwartzman\inst{4}
  \and Hubert Wagner\inst{3}
  \and Rien van de Weygaert\inst{5}
}

\institute{Univ Lyon, ENS de Lyon, Univ Lyon1, CNRS, Centre de Recherche Astrophysique de Lyon UMR5574, F--69007, Lyon, France
  \and Technion - Israel Institute of Technology, 32000, Haifa, Israel
  \and IST Austria (Institute of Science and Technology Austria), 3400 Klosterneuburg, Austria
  \and Division of Biostatistics, University of California, San Diego, California, USA
  \and Kapteyn Astronomical Institute, Landleven 12, 9747 AG Groningen, the Netherlands
}


\abstract{
We study the topology generated by the temperature fluctuations of the Cosmic
Microwave Background (CMB) radiation,
as quantified by the number of components and holes, formally given by the Betti numbers, in the growing excursion sets.
We compare CMB maps observed by the Planck satellite with a thousand simulated maps generated according to the LCDM paradigm with Gaussian distributed fluctuations. The comparison is multi-scale, being performed on a sequence of degraded maps with mean pixel separation ranging from 0.05 to 7.33 degrees. 

The survey of the CMB over $\Sspace^2$ is incomplete due to obfuscation effects by bright point sources and other extended foreground objects like our own galaxy. To deal with such situations, where analysis in the presence of "masks" is of importance, we introduce the concept of relative homology.

The parametric $\chi^2$-test shows differences between observations and simulations, yielding $p$-values at per-cent to less than per-mil levels roughly between 2 to 7 degrees, with the difference in the number of components and holes peaking at more than $3\sigma$ sporadically at these scales. The highest observed deviation between the observations and simulations for $\relBetti{0}$ and $\relBetti{1}$ is approximately between $3\sigma$-4$\sigma$ at scales of 3 to 7 degrees.  There are reports of mildly unusual behaviour of the Euler characteristic at 3.66 degrees in the literature, computed from independent measurements of the CMB temperature fluctuations by Planck's predecessor WMAP (Wilkinson Microwave Anisotropy Probe) satellite. The mildly anomalous behaviour of Euler characteristic is phenomenologically related to the strongly anomalous behaviour of components and holes, or the zeroth and the first Betti numbers, respectively. Further, since these topological descriptors show anomalous behaviour consistently over independent measurements of Planck and WMAP, instrumental errors and systematics may be an unlikely source. These are also the scales at which the observed maps exhibit low variance compared to the simulations, and approximately the range of scales at which the power spectrum exhibits a dip with respect to the theoretical model. Non-parametric tests show even stronger differences at almost all scales. Crucially, Gaussian simulations, based on power spectrum matching the characteristics of the observed dipped power spectrum, are not able to resolve the anomaly. Understanding the origin of the anomalies in the CMB -- whether cosmological in nature, or arising due to late-time effects is an extremely challenging task. Regardless, beyond the trivial possibility that this may still be a manifestation of an extreme Gaussian case, these observations, along with the super-horizon scales involved, may motivate to look at primordial non-Gaussianity.  Alternative scenarios worth exploring may be models with non-trivial topology.}


\maketitle


\section{Introduction}
\label{sec:intro}

\emph{Cosmological background.}
The Lambda Cold Dark Matter (or LCDM) standard paradigm of cosmology postulates that the Universe consists primarily of cold non-relativistic \emph{dark matter}, which reveals its presence only through gravitational interactions, and the Universe is currently driven by \emph{dark energy}, causing accelerated volume expansion in this model. The \emph{Cosmic Microwave Background} (or CMB) \emph{radiation}, which originates at the epoch of recombination, is the most important observational probe into the validity of the standard paradigm today \citep{jones2017precision}. 
It is the earliest visible light and offers a glimpse into the processes during the nascent stage of the Universe. Fluctuations about the mean in the temperature field of the CMB correspond to the fluctuations in the distribution of matter in the early Universe, thus understanding the  CMB is crucial to understanding the primordial Universe.

The LCDM paradigm together with the inflationary theories in their simplest forms, predict the primordial
perturbations to be realizations of a homogeneous and isotropic
Gaussian random field \citep{GuthPi1982}.  This hypothesis is supported  experimentally by  CMB observations 
\citep{smoot1992,bennett2003,Spergel2007,Komatsu2011,Planck2015cosmoparams}
and theoretically  by the Central Limit Theorem.  While it has largely been agreed upon that the CMB exhibits characteristics of a homogeneous and isotropic Gaussian field, there are lingering doubts.  The pioneering works of \cite{eriksen04} and \cite{Park2004} challenge the assumption of homogeneity, and the alignment of low multipoles \citep{multipoles} challenges the assumption of isotropy.  Other noted anomalies include the vanishing correlation function at large scales, and the unusually low variance at approximately $3$ degrees; see \cite{CMBAnomaliesStarkman} for a review and possible interpretations. \cite{PlanckXXIII} independently confirms these anomalies.

The primordial non-Gaussianity remains a topic of ongoing debate. Deviations from Gaussianity, if found, will point to new physics driving the Universe in its nascent stages. The consensus is biased towards its absence \citep{Komatsu2011,PlanckXXIII,matsubara2010,bartolo10,Planck2013}; see also \citep{BFS17} for a review and a model-independent route of analysis. 
Despite mildly unusual behaviour of the Euler characteristic,
pointed out in \cite{Eriksen04NG} and \cite{Park2004},
the methods employed until today have not provided compelling evidence
of non-Gaussianity in the CMB. 
In contrast, the topological methods of this paper find the observed CMB maps \citep{planckCompSepMaps} to be significantly different from the  Full Focal Plane 8 (FFP8) simulations \citep{PlanckXXIII,planckSims} that assume the initial perturbations to be Gaussian.

\medskip\noindent\emph{Topological methods.}
Topology is the branch of mathematics concerned with  properties of shapes and spaces
preserved under continuous deformations, such as stretching and bending, but not tearing and gluing. It is related to, but different from geometry, which measures size and shape. Both geometry and topology have been used in the past 
to study the structure of the CMB radiation and other cosmic fields.
Historically, the predominant tools in this endeavour
were the \emph{Minkowski functionals},
which for a $2$-manifold embedded in the $3$-dimensional space are related to the enclosed volume, the area, the total mean curvature, and the total Gaussian curvature.
By the Gauss--Bonnet Theorem, for $2$-manifolds, the latter is $2 \pi$ times the \emph{Euler characteristic} \citep{Eul58}, thus providing a bridge between geometry and topology.
Early topological studies of the cosmic mass distribution were based on the
Euler characteristic of the iso-density surfaces, which generically are $2$-manifolds
\citep{Dor70,bbks,GDM86,PPC13}.
The full set of Minkowski functionals was later introduced to cosmology in \citep{Mecke94,schmalzing1997,schmalzinggorski}. 
For Gaussian, and Gaussian-related random fields, the expected values of the Minkowski functionals
of excursion sets have known analytic expressions \citep{Adl81,Adl10}, which is one of the main 
reasons they have played a key role in the study of real valued fields arising in cosmology and other disciplines.

While the Minkowski functionals have been instructive, the topological information
contained in them is limited and convolved with geometric information. 
Moreover, they are not equipped to address the hierarchical aspects of the matter distribution directly, although partial Minkowski functionals \citep{partialMF} may be useful in certain settings.  We, therefore, analyse CMB  fluctuations in terms of the purely topological concepts of \emph{homology} \citep{munkres1984elements}, as quantified by \emph{Betti numbers} \citep{Bet71} and \emph{persistence} \citep{ELZ02,EdHa10}.  
By the Euler-Poincar\'{e} formula, the Euler characteristic is the alternating sum of the Betti numbers, implying that the latter provide a finer description of topology \citep{munkres1984elements}. 
A broad exposition of these concepts, in a cosmological setting,
is given in  \citep{PEW16,pranavThesis,ISVD10}; also see  \citep{PPC13,Sousbie1,Shivashankar2015,RST,Makarenko18,CS18} for some applications. Related but slightly different methodologies used for the analysis of cosmological datasets, emanating from concepts in Morse theory maybe found in \citep{colombi1,colombi2,colombi3}.

\medskip\noindent\emph{Results.}
Our main result is an anomaly of the observed CMB radiation when compared with simulations based on Gaussian prescriptions.  The $\chi^2$-test yields a significant difference between the number of components and holes in the observed sky compared to the simulations, with $p$-values at per-cent to less than per-mil levels at scales of roughly 2 to 7 degrees. The differences peak sporadically at more than $3\sigma$ at these scales.  Non-parametric tests see an even more significant difference between the observation and the simulations at almost all scales. The $\chi^2$-test shows the anomaly at roughly the same scales at which the power spectrum exhibits a dip. \cite{Eriksen04NG} reports a mildly unusual Euler characteristic at approximately 3 degrees in the earlier measurements of the CMB radiation by the Wilkinson Microwave Anisotropy Probe (WMAP) satellite, which is related to the anomalous behaviour of components and holes.
The noted anomaly motivates a closer look at the standard paradigm.  Possible scenarios include, but are not limited to, primordial non-Gaussianity, as well as models with non-trivial topology \citep{SteinerHyperbolic,SteinerMultiConnected,bernui18}.

\medskip\noindent\emph{Overview.} The workflow in this paper is straightforward:
topological descriptors are computed from cosmology data,
and statistical tests based on these descriptors
are used to compare the observations with simulations. Seection~\ref{sec:22} gives a summary of the topological concepts. Section~\ref{sec:data} describes the data, the computational pipeline, and a brief account of the statistical tests employed. Section~\ref{sec:3} presents the main results of the paper, followed by a summary and conclusions in Section~\ref{sec:discussion}.

\section{Topological Background}
\label{sec:22}

Since the CMB radiation is observed as
a scalar field on the $2$-dimensional sphere,
the topological concepts needed in this paper are elementary,
namely the components and the holes of subsets of this sphere.
To count them in the presence of regions with unreliable data,
we compute the ranks of the homology groups relative to the mask
that covers these regions.

\subsection{Excursion sets and absolute homology}
\begin{figure}
	\centering
	\includegraphics[width=0.49\textwidth]{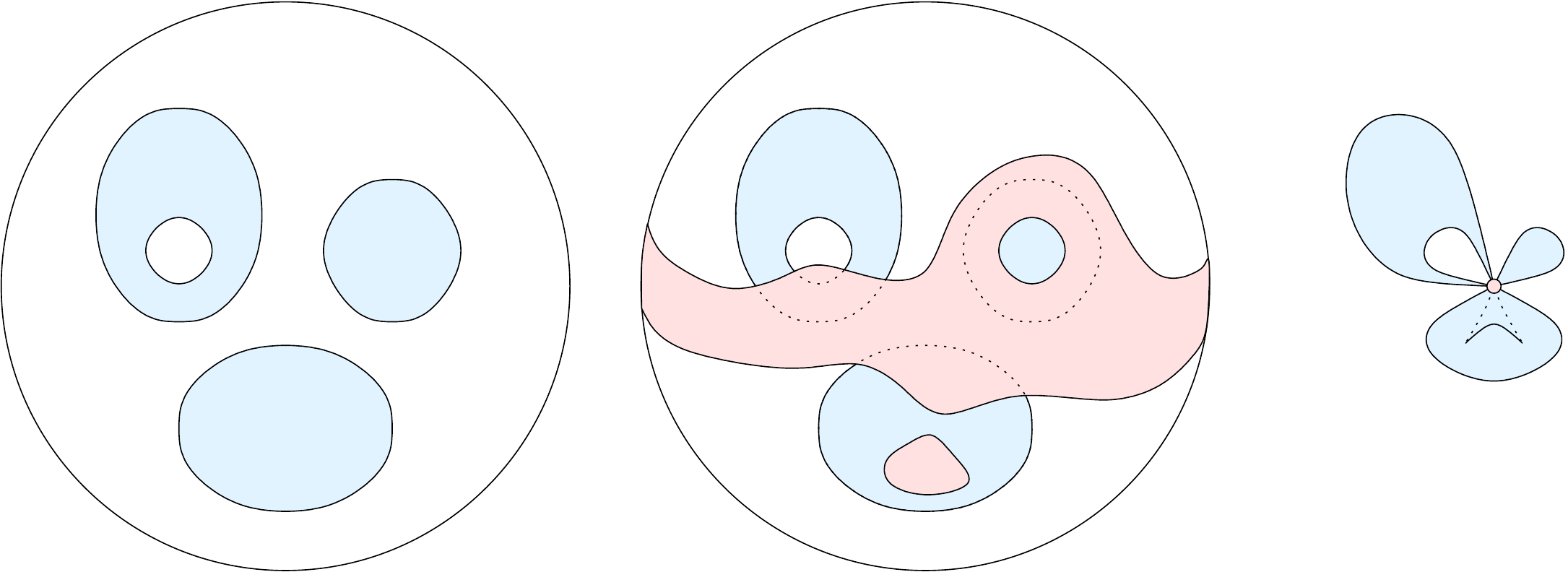} 
	\caption{\emph{Left}: A blue excursion set on the sphere consisting of an upper left component with a hole, an upper right component, and a lower component. Its Betti numbers are $\Betti{0} = 3$, $\Betti{1} = 1$, $\Betti{2} = 0$, and its Euler characteristic is $\Euler = 3-1+0 = 2$. \emph{Middle}: A pink mask in which the data is not reliable. It covers part of the upper left component and hole, its hole is fully contained in the upper right component, and it overlaps the lower component in two disconnected pieces. \emph{Right}:  A visualization of the relative homology groups obtained by shrinking the mask to a point and pulling the excursion set with it. We have $\relBetti{0} = 0$ because all three components connect to the shrunken mask, $\relBetti{1} = 2$ because the loop in the upper left component is preserved and a new loop in the lower component is formed, and $\relBetti{2} = 1$ because the upper right component takes on the shape of sphere. The (relative) Euler characteristic is therefore $\relEuler = 0-2+1 = -1$.}
	\label{fig:AbsRel}
\end{figure}

Writing $\Sspace^2$ for the $2$-dimensional sphere
and $f \colon \Sspace^2 \to \Rspace$ for the temperature field of the CMB,
the \emph{excursion set} at a temperature $\nu$ is the subset of the sphere
in which the temperature is $\nu$ or larger:
$\Excursion (\nu)  =  \{ x \in \Sspace^2  \mid  f(x) \geq \nu \}$.
It is a closed set, and we write
$\Betti{0} (\nu)$ for its number of \emph{components}.
A \emph{hole} is a component of the complement,
$\Sspace^2 \setminus \Excursion (\nu)$.
Assuming there is at least one hole, we write
$\Betti{1} (\nu) + 1$ for the number of holes,
and we set
 $\Betti{2} (\nu) = 0$, because $\Excursion (\nu)$ does not
cover the entire sphere.
On the other hand, if there is no hole, we set
$\Betti{1} (\nu) = 0$ and $\Betti{2} (\nu) = 1$;
see the left panel of Figure~\ref{fig:AbsRel} for an illustration.
These definitions are motivated by the more general theory
\citep{munkres1984elements}
in which the \emph{$p$-th Betti number} is the rank of the
\emph{$p$-th homology group}:
$\Betti{p} = \Rank{\Homology{p}}$ for $p = 0, 1, 2$. These are the basic objects of homology.
The \emph{Euler characteristic} of the excursion set is the alternating sum
of the Betti numbers:
$\Euler (\nu) = \Betti{0} (\nu) - \Betti{1} (\nu) + \Betti{2} (\nu)$.

The Euler characteristic has a long history in the CMB literature,
largely due to the fact that a simple analytic formula
for its expected value is known when the CMB is modeled as a Gaussian random field
\citep{Adl81,Adl10}.
While such formulas are not known for the Betti numbers,
the information they carry is richer. \cite{grfPranav} present a numerical study of the Betti numbers of Gaussian random fields, and compare them to the Euler characteristic and Minkowski functionals, and find that Betti numbers present a more detailed account of the topological properties of the field compared to the Euler characteristic (also see \cite{PPC13}).
In general, near the mean level of $\nu$, one expects the components
and holes of the excursion set to be of similar size and number.
Accordingly, one expects $\Betti{0} (\nu)$ and $\Betti{1} (\nu)$ to
be of similar magnitude,
combining to give an Euler characteristic close to zero.
Such an Euler characteristic tells us nothing about the individual Betti numbers, beyond the fact that they are similar.

\subsection{Masks and relative homology}
We define the \emph{mask} to be the region in which the data is not reliable,
and denote it by $\Mask \subseteq \Sspace^2$.
In our application, it includes a belt around the equator corresponding to 
the thickened disk of the Milky Way, along with other galactic and extra-galactic bright foreground objects that
interfere with the observation of the CMB radiation.
In an effort to exclude the mask from our computations,
we consider the \emph{reduced excursion set}: $\Excursion (\nu) \setminus \Mask$.
Treating $\Mask$ as a closed set, this difference is not necessarily closed.
An appropriate topological measure is the
\emph{relative homology} of a pair of closed spaces, $(E, M)$,
with the second being contained in the first.
In our setting, the pair is $E = \Excursion (\nu)$
and $M = \Mask \cap \Excursion (\nu)$.
Just as in the absolute case, we get relative homology groups in
dimensions $0$, $1$, and $2$, and we use their ranks for quantification.
It is tempting to refer to these ranks as \emph{relative Betti numbers},
but this is not the traditional terminology, and we just write
$\relBetti{p} = \Rank{\Homology{p} (E, M)}$ for $p = 0, 1, 2$.
If $M = \emptyset$, then $\relBetti{p} = \Betti{p}$,
for all three choices of $p$,
but if the mask overlaps with the excursion set, then there are differences.
We explain some of these differences with reference to Figure~\ref{fig:AbsRel}:
If $\Mask$ overlaps a component of $E$, this component is no longer counted
because every vertex in it bounds a path connecting it to the mask.
If $\Mask$ overlaps the component in two disconnected pieces,
we count a new loop, namely the path connecting these two pieces.
If $\Mask$ covers part of a hole, this hole is still counted
because the part of its boundary curve outside the mask is open,
with endpoints in $\Mask$.
If a hole of $\Mask$ is contained in the excursion set, we get a surface
without boundary. The relation between absolute and relative homology is compactly
expressed by the \emph{exact sequence of the pair} $M \subseteq E$ \citep{munkres1984elements}:

\begin{align}
0  &\to  \Homology{2}(M)  \to  \Homology{2}(E)  \to  \Homology{2}(E,M)          
\to  \Homology{1}(M)  \to  \Homology{1}(E) \\ \nonumber
&\to  \Homology{1}(E,M)     \to  \Homology{0}(M)  \to  \Homology{0}(E)  \to  \Homology{0}(E,M)  \to  0;
\end{align}

\noindent Without going into details, this means that we can assign non-negative integers
to the arrows so that the rank of each group is the sum of integers assigned to its incoming
and outgoing arrows.
For example in Figure~\ref{fig:AbsRel}, we have
$0 \to 0 \to 0 \to 1 \to 1 \to 1 \to 2 \to 4 \to 3 \to 0 \to 0$,
and it is easy to find the assignment of integers that
satisfies the stated property.

\begin{figure}
	\centering
	\subfloat{\includegraphics[width=0.5\textwidth]{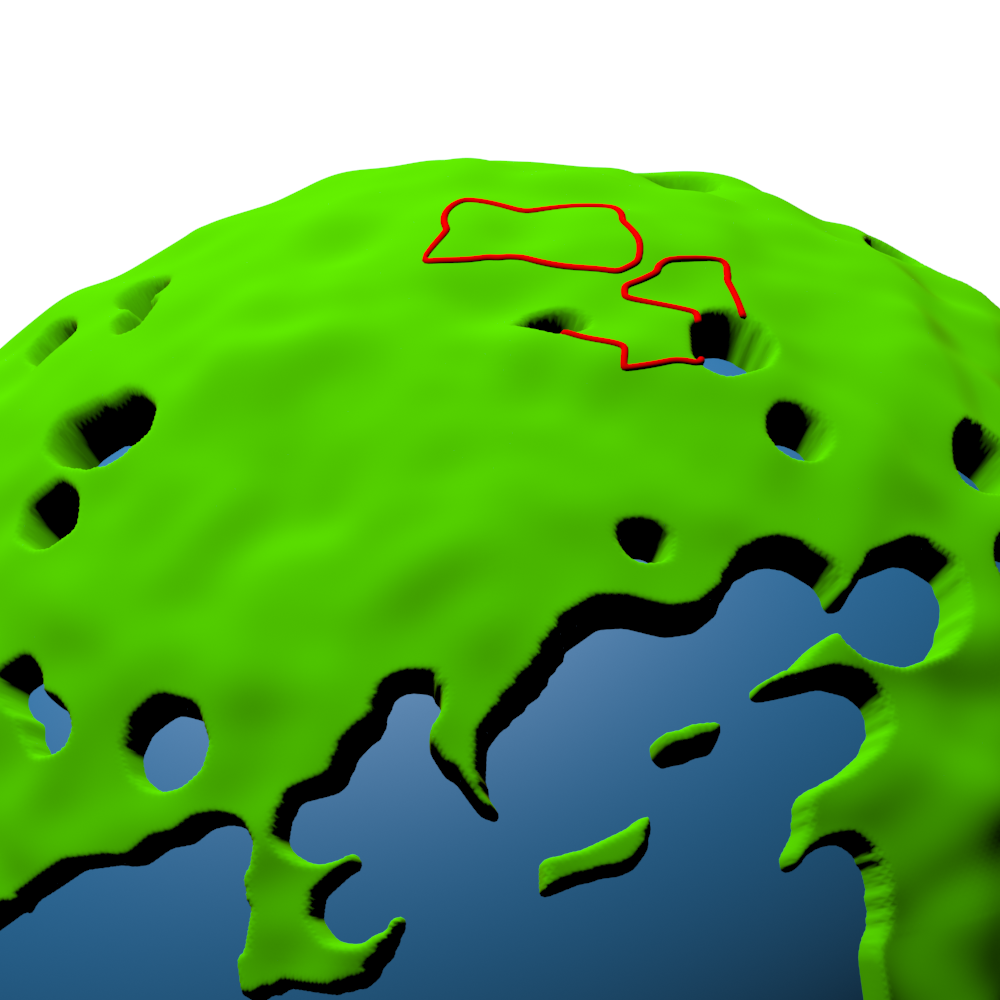} }
	\caption{A small section of the sphere of directions, with the
temperature field visualized by the green landscape that complements
the blue mask drawn at lower altitude.  We see one closed loop, surrounding a relative depression of the temperature field, and two open loops, connecting points in the mask along locally highest paths. The visualization is based on the observed CMB maps cleaned using the NILC technique, and smoothed at 4 degrees.}
	\label{fig:loop}
\end{figure}

\begin{figure*}
	\centering     \resizebox{!}{1.5in}{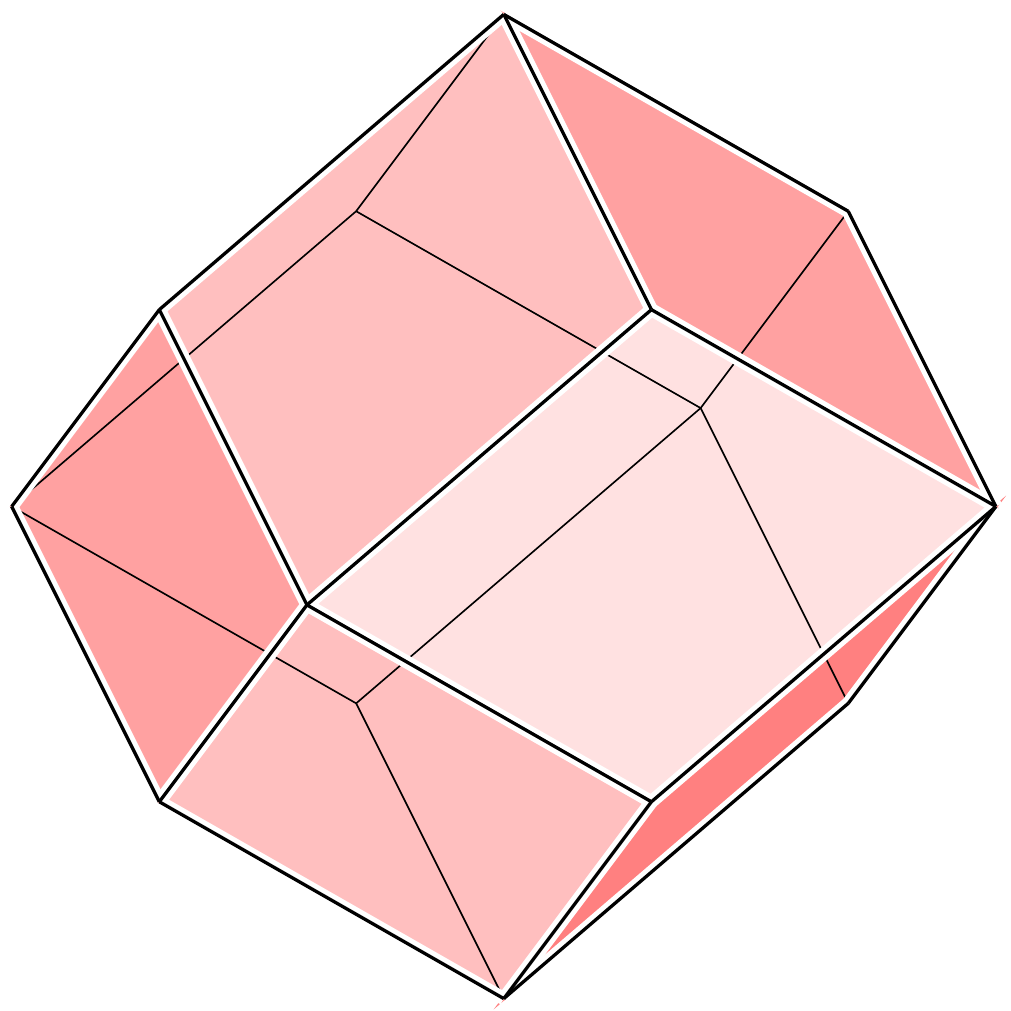}
	\hspace{0.1in} \resizebox{!}{1.5in}{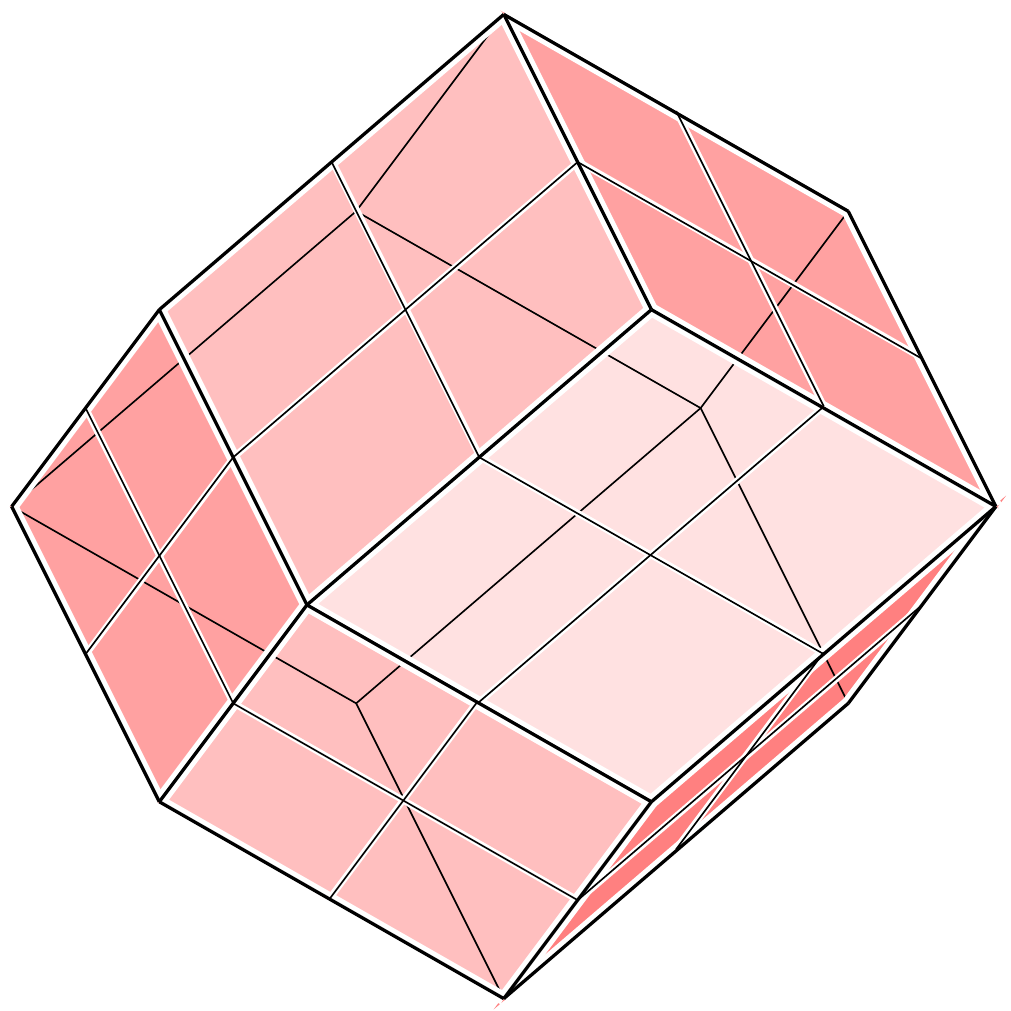}
	\resizebox{!}{1.5in}{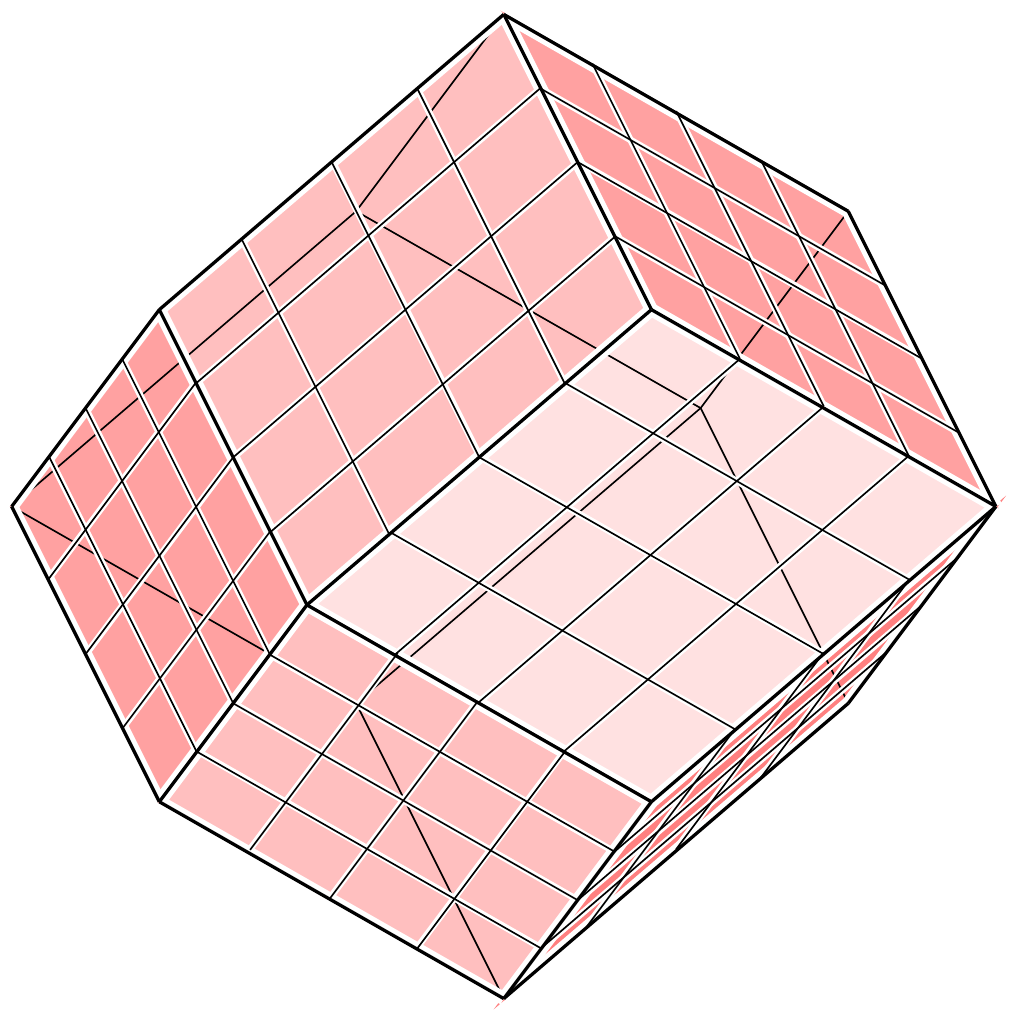}
	\hspace{0.1in} \resizebox{!}{1.5in}{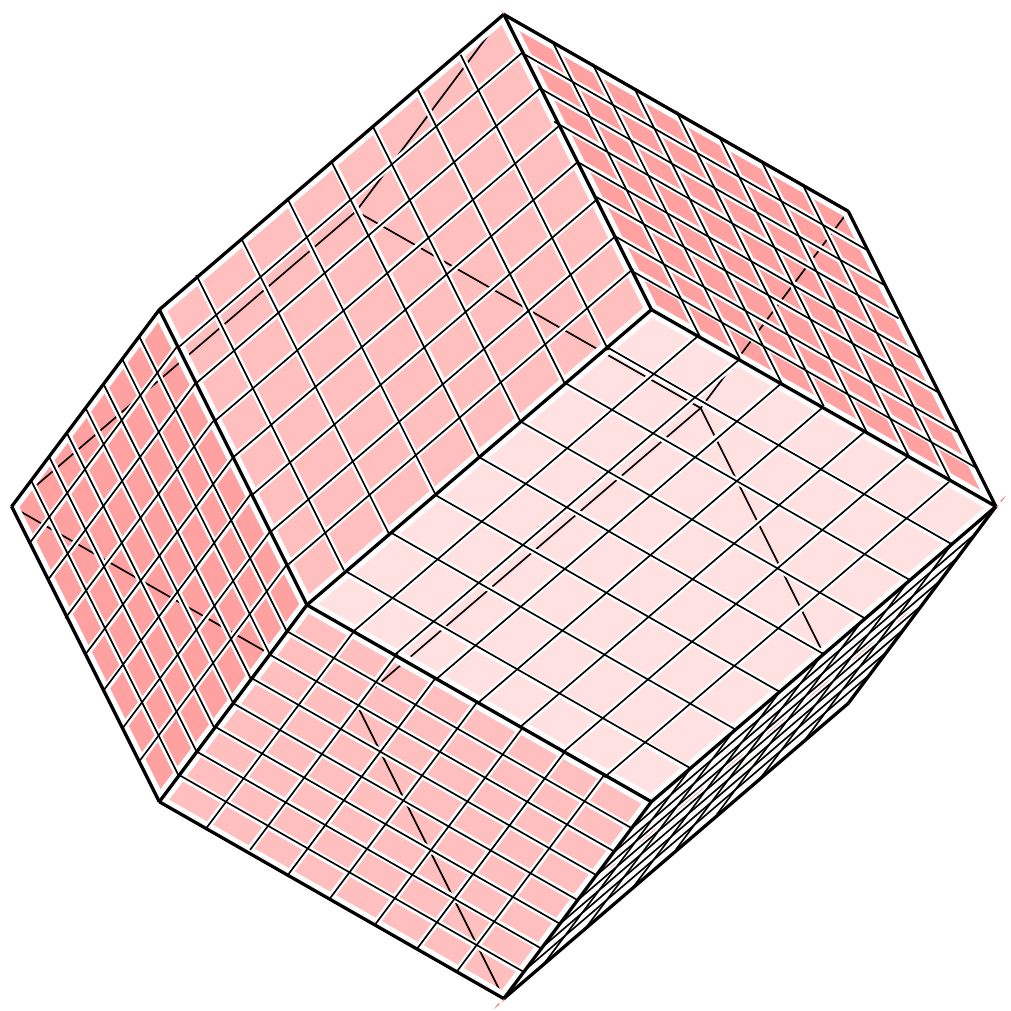}
	\caption{The facets of the rhombic dodecahedron serve as patches in
		the HealPix representation of the sphere.
		In sequence, we show the $12$ patches decomposed into
		$1, 4, 16,64$ pixels.
		The final representation is obtained by central projection of the
		pixel centers and a distortion yielding an approximately
		equal-area decomposition of the sphere (not shown).}
	\label{fig:healpix}
\end{figure*}

\subsection{Variationally maximal loops}
When we count $\Betti{1}+1$ holes in absolute homology,
we really count $\Betti{1}$ \emph{loops} needed to
separate them. In relative homology, the connection is not as intuitive because we also
have open loops, whose endpoints lie in the mask; see Figure~\ref{fig:loop}.
Generally, there are uncountably many ways to draw a loop,
and in homology they are all considered equivalent.
The set of equivalent loops is called a \emph{homology class},
and any one of the loops in the class is a
\emph{representative}.
These classes are the elements of the $1$-dimensional homology group,
which is a vector space. The rank of this group counts the classes that are needed to span the vector space.

For visualization, it is desirable to have a unique representative for each class.
Similar to the intuitive notion of the rim of a crater,
we choose this representative as high as possible, alternating between peaks and saddles
of $f$ which it connects via ridges within the reduced excursion set.
We refer to this loop as the \emph{variationally maximal representative}
of its class; see Figure~\ref{fig:loop} for an example.
While constructing variational maxima for smooth scalar fields may be problematic,
the persistence algorithm applied to a piecewise linear scalar field
produces them as a side effect of reducing the boundary matrix;
see \ref{sec:computation}.

\begin{figure}
	\centering   
	\includegraphics[width=0.5\textwidth]{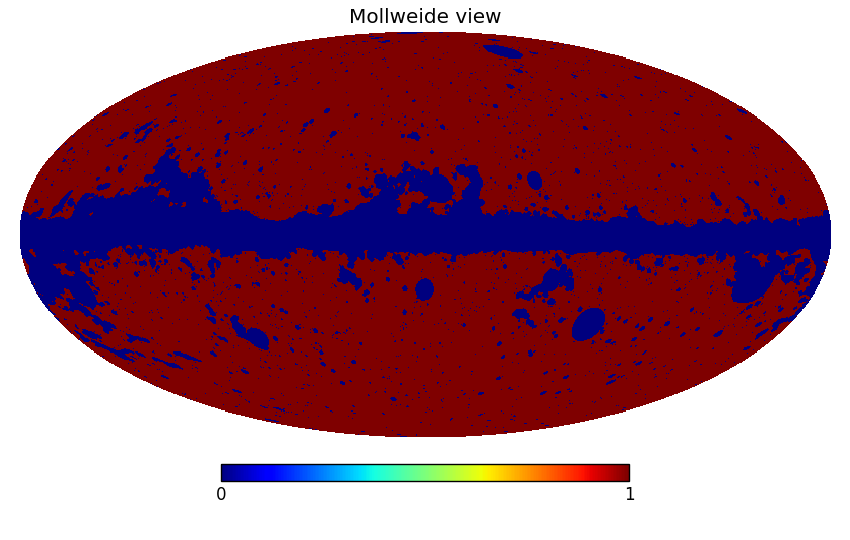}
	\caption{The UT78 mask released by the Planck team. It is a conservative mask, that masks the known point sources and other bright foreground objects, in addition to the galactic disk.}
	\label{fig:UT78}
\end{figure}

\begin{figure*}
	\centering   
	\subfloat[]{\includegraphics[width=0.24\textwidth]{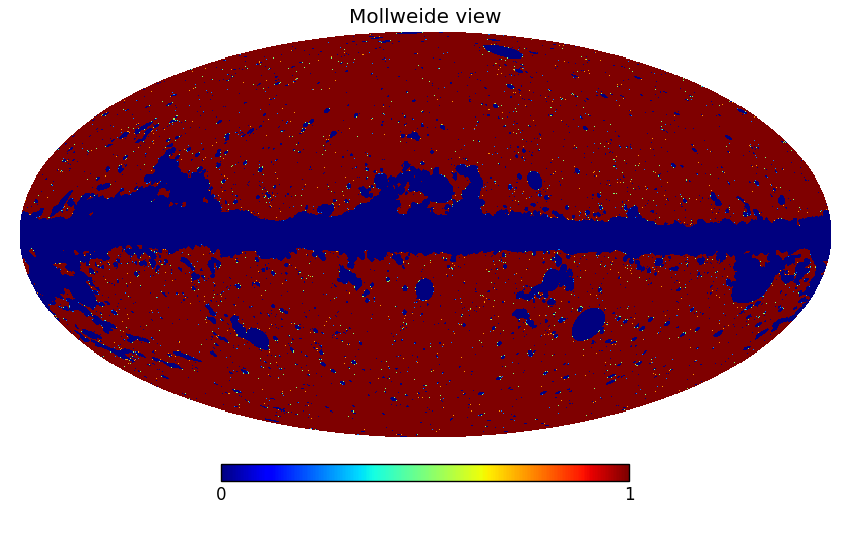}}
	\subfloat[]{\includegraphics[width=0.24\textwidth]{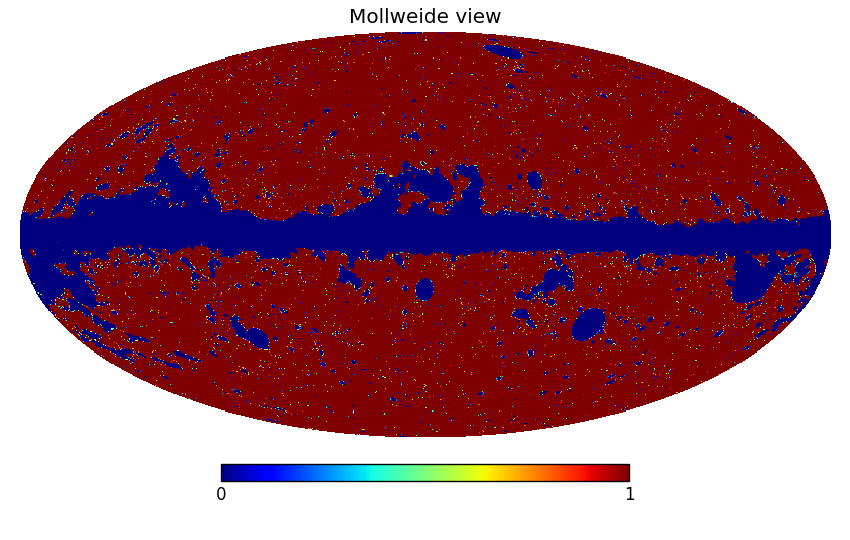}}
	\subfloat[]{\includegraphics[width=0.24\textwidth]{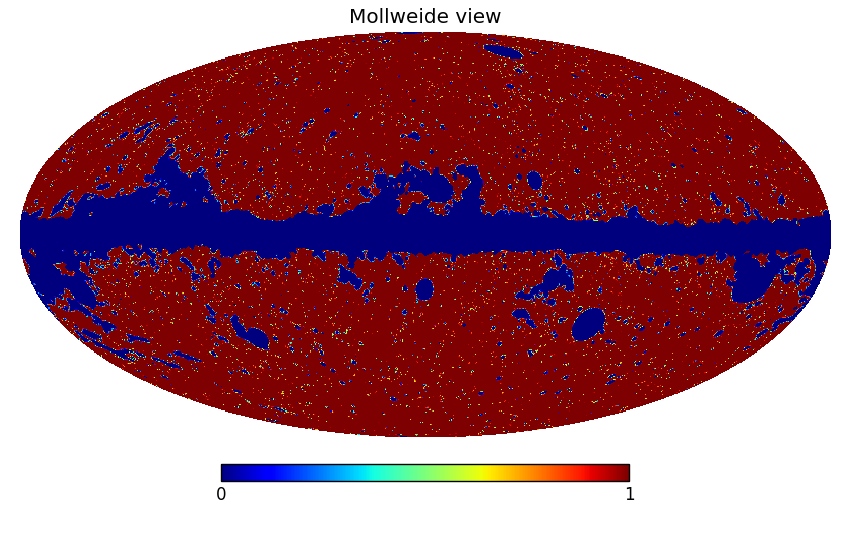}}
	\subfloat[]{\includegraphics[width=0.24\textwidth]{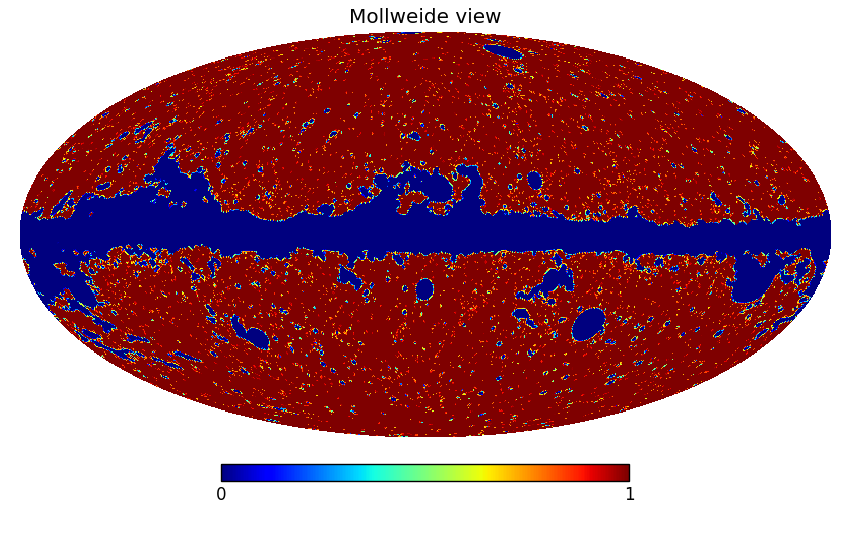}}\\
	\subfloat[]{\includegraphics[width=0.24\textwidth]{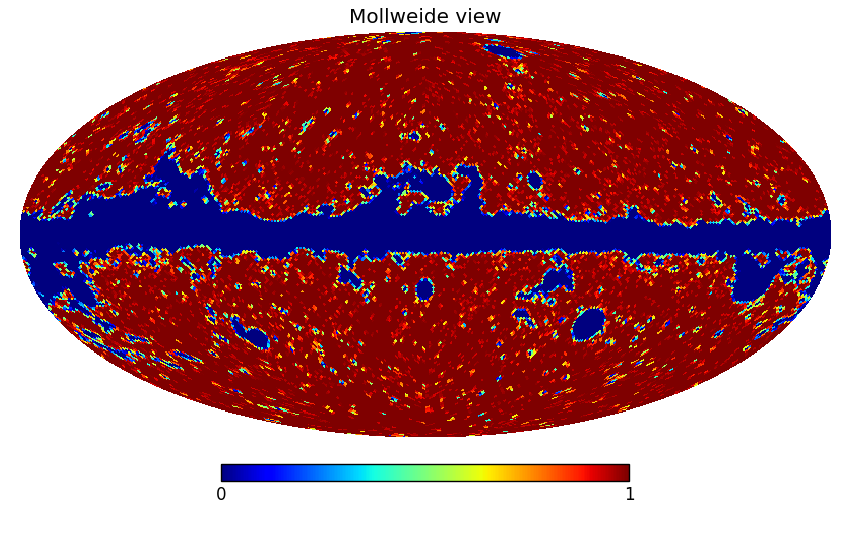}}
	\subfloat[]{\includegraphics[width=0.24\textwidth]{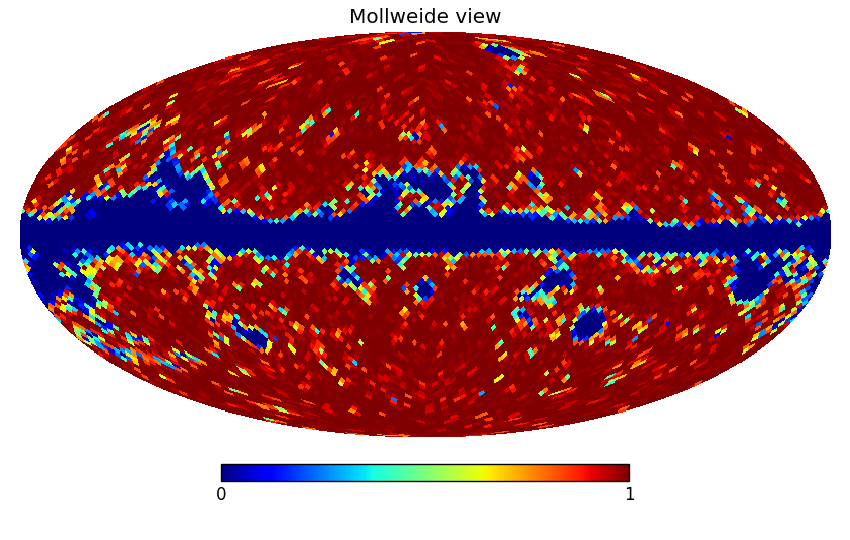}}
	\subfloat[]{\includegraphics[width=0.24\textwidth]{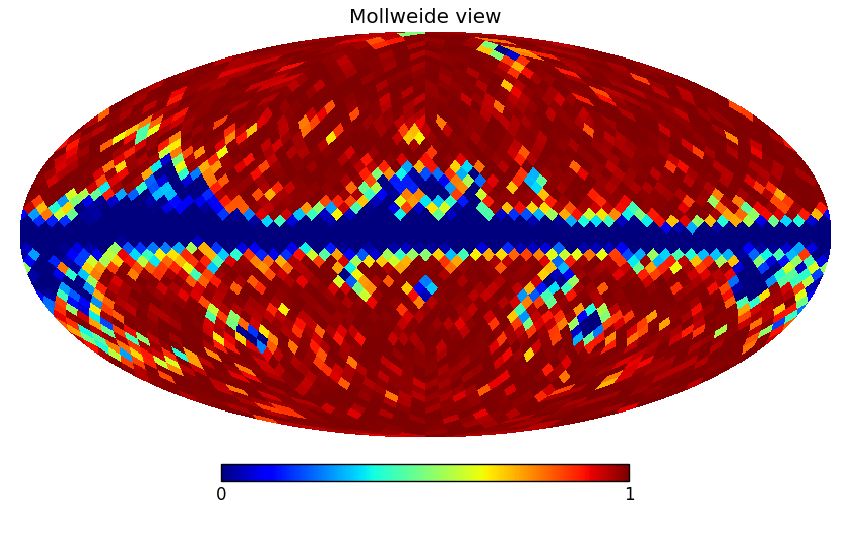}}
	\caption{The degraded masks before binarization. For high enough resolutions, the masks have similar appearance as the original one, but are distinguishable when zoomed in to.}
	\label{fig:degradedNBMask}
\end{figure*}

\begin{figure*}
	\centering   
	\subfloat[]{\includegraphics[width=0.24\textwidth]{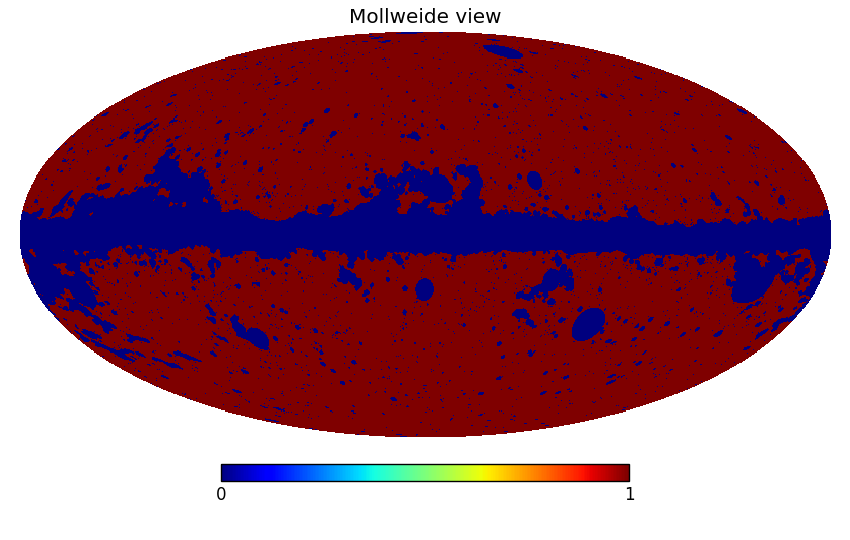}}
	\subfloat[]{\includegraphics[width=0.24\textwidth]{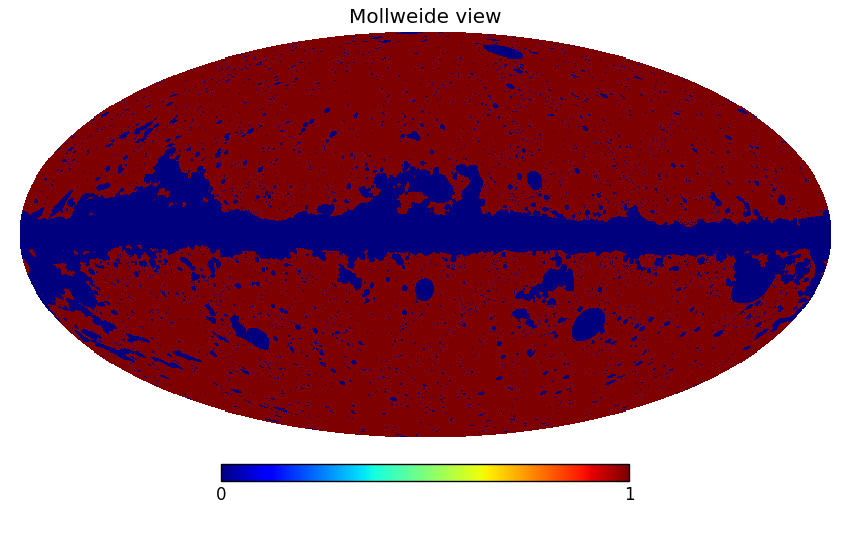}}
	\subfloat[]{\includegraphics[width=0.24\textwidth]{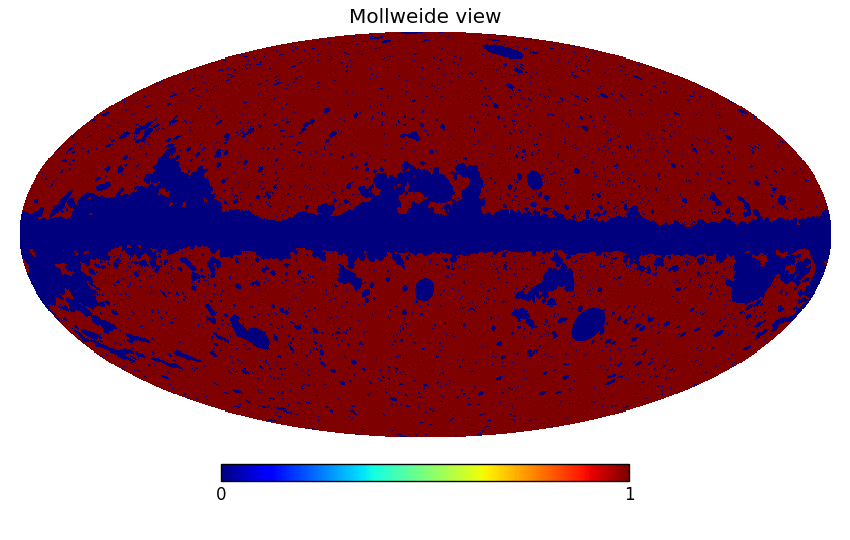}}
	\subfloat[]{\includegraphics[width=0.24\textwidth]{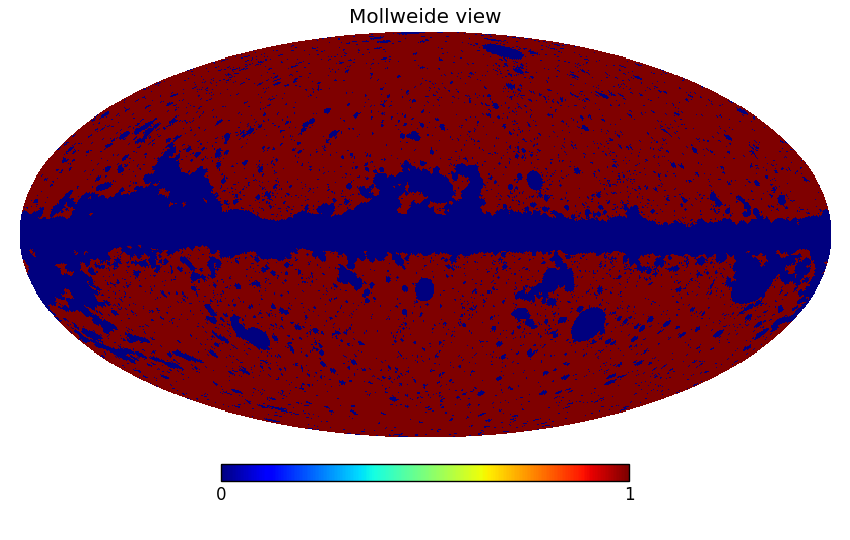}}\\
	\subfloat[]{\includegraphics[width=0.24\textwidth]{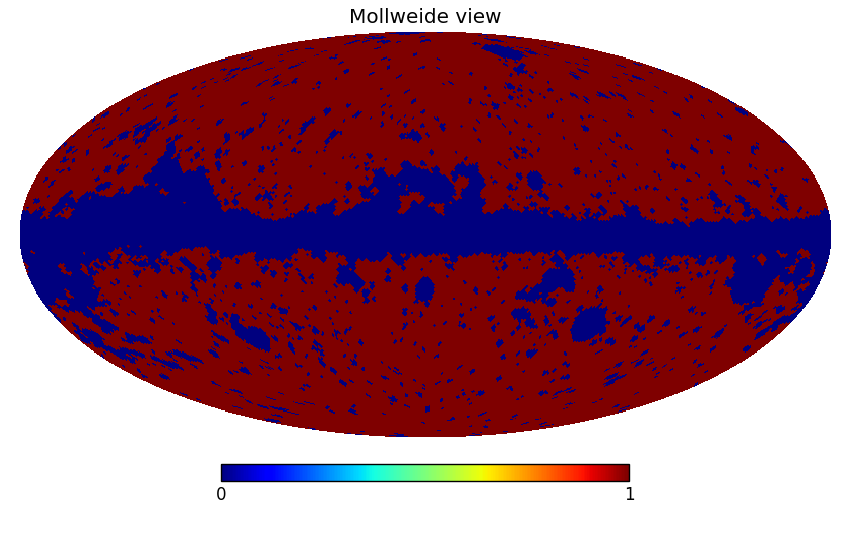}}
	\subfloat[]{\includegraphics[width=0.24\textwidth]{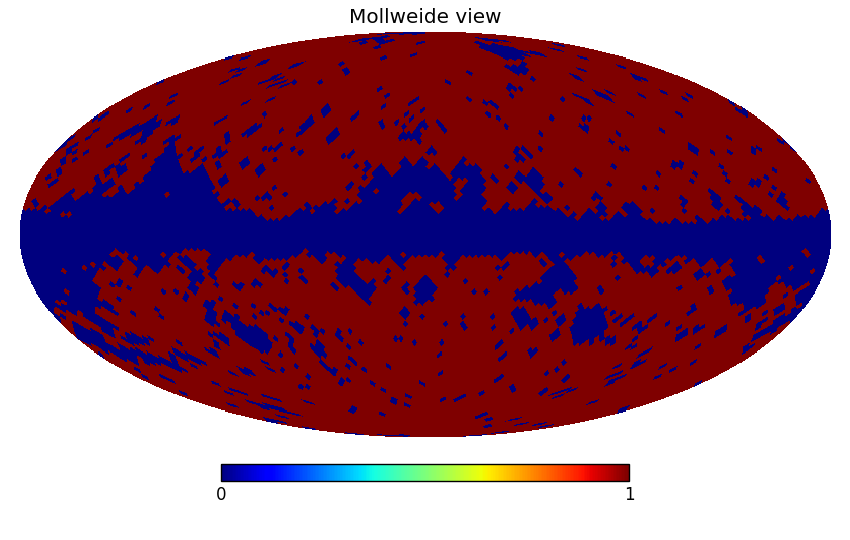}}
	\subfloat[]{\includegraphics[width=0.24\textwidth]{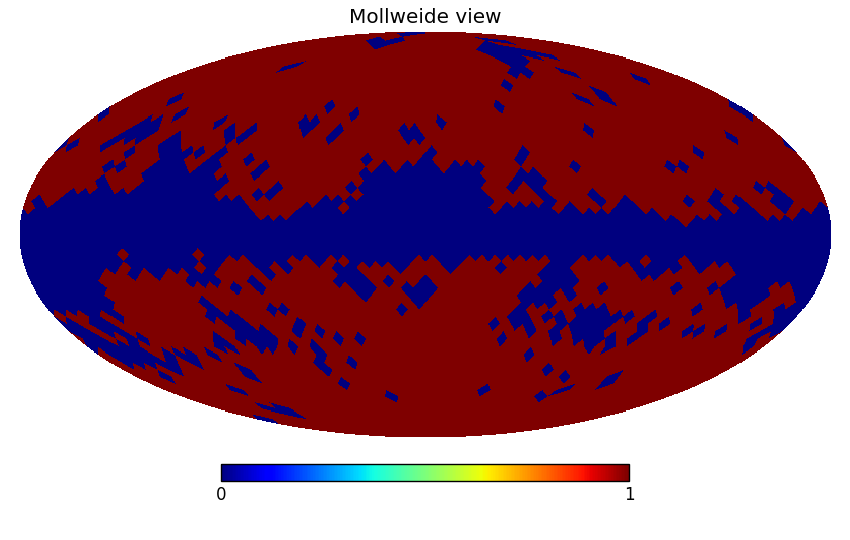}}
	
	\caption{The degraded masks after binarization, thresholded at $0.9$.}
	\label{fig:degradedBMask}
\end{figure*}


\begin{figure*}
	\centering   
	\subfloat[]{\includegraphics[width=0.24\textwidth]{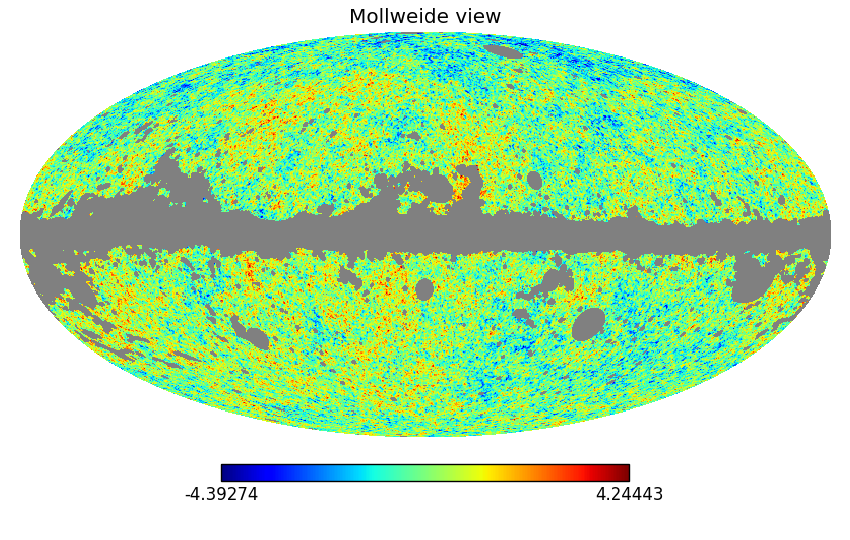}}
	\subfloat[]{\includegraphics[width=0.24\textwidth]{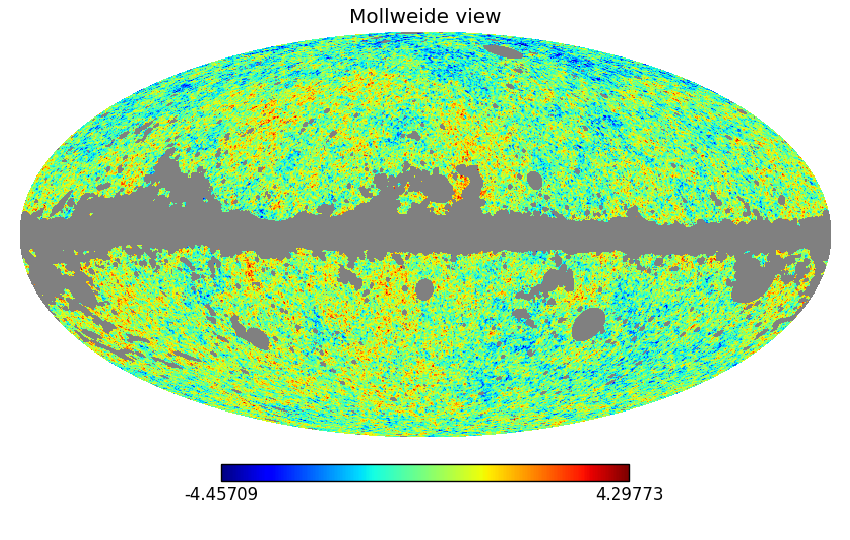}}
	\subfloat[]{\includegraphics[width=0.24\textwidth]{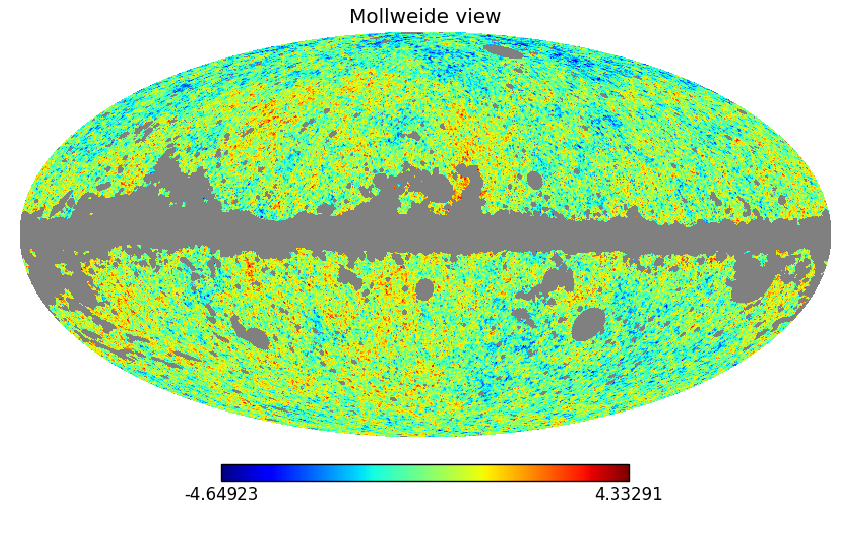}}
	\subfloat[]{\includegraphics[width=0.24\textwidth]{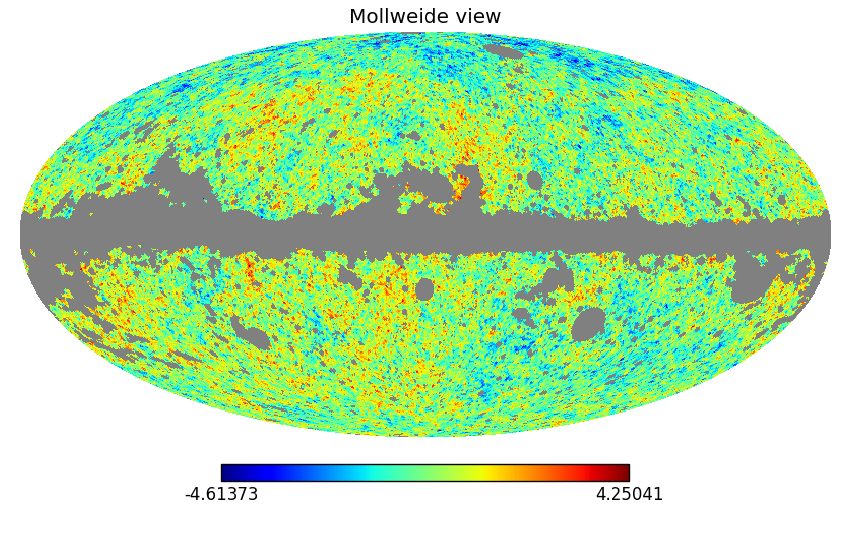}}\\
	\subfloat[]{\includegraphics[width=0.24\textwidth]{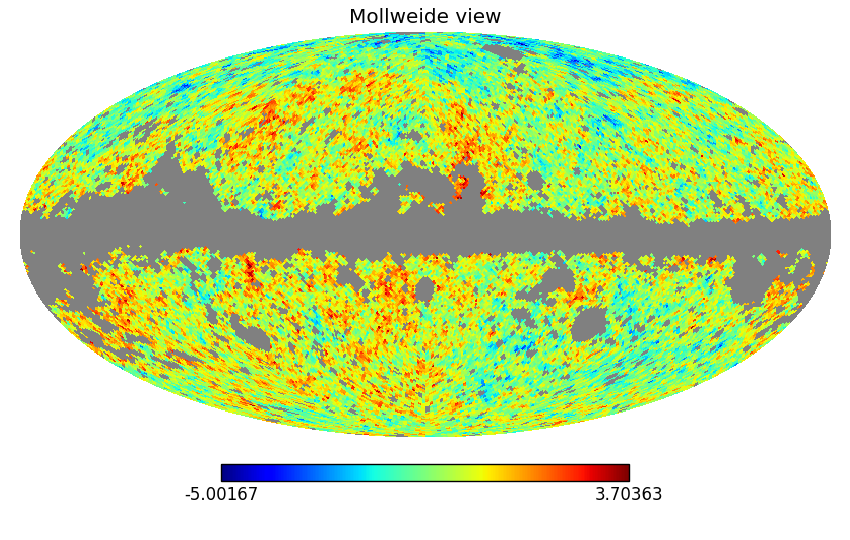}}
	\subfloat[]{\includegraphics[width=0.24\textwidth]{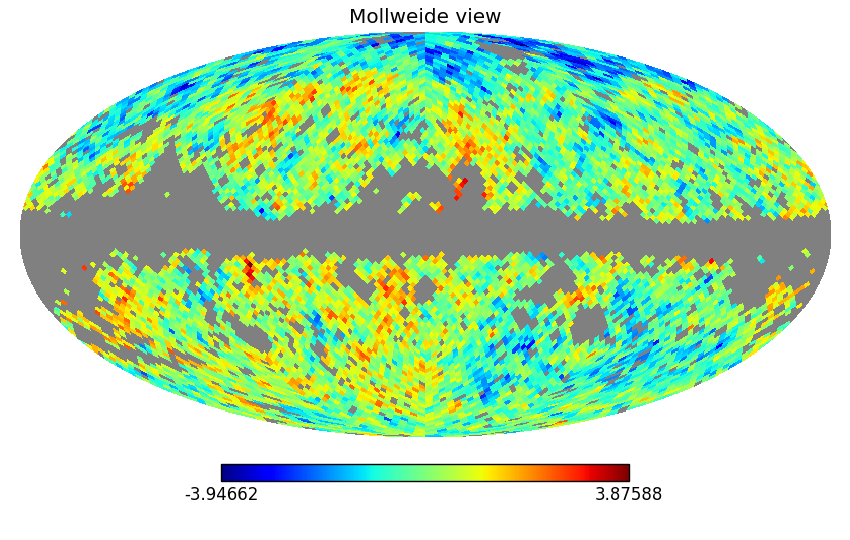}}
	\subfloat[]{\includegraphics[width=0.24\textwidth]{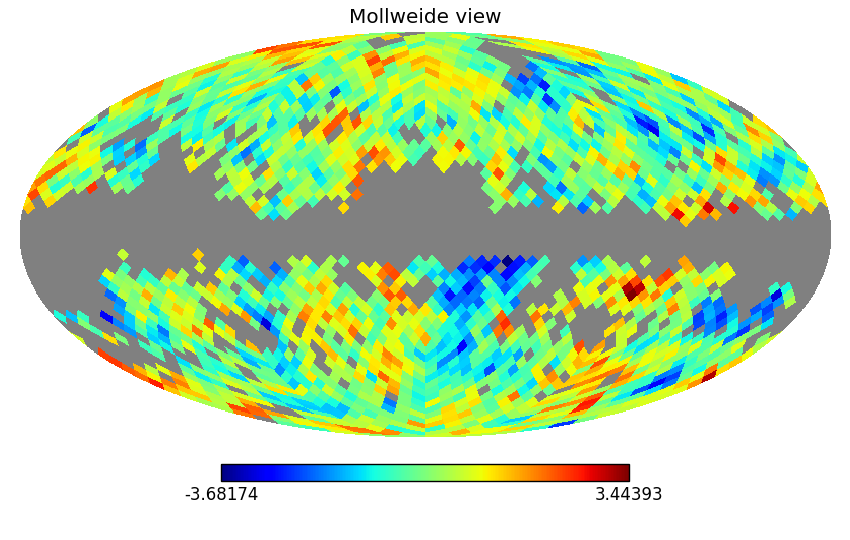}}
	\caption{A visualization of the masked maps at various degraded levels.}
	\label{fig:degradedMaskedMap}
\end{figure*}
\section{Data and methods}
\label{sec:data}

\subsection{Data}

Decades after the accidental discovery of the CMB, its first space based observational probe was carried out by
the Cosmic Microwave Background Explorer (COBE) satellite \citep{smoot1992}, 
establishing that the CMB is perfect black-body radiation.
Later, the Wilkinson Microwave Anisotropy Probe was launched
to study the temperature fluctuations in greater detail \citep{Spergel2007}.
Most recently, the high precision Planck mission was launched,
measuring  temperature fluctuations to an accuracy
of $10^{-5}$ degrees \citep{planckMain},
and  at a resolution of
five arc-minutes, giving  the most detailed and precise measurement
of  CMB temperature fluctuations  currently available. We use the Planck maps for our analyses \citep{planckCompSepMaps,planckSims}.

\medskip\noindent\emph{Format.}
The CMB sky maps are presented in the \texttt{HealPix} format \citep{Healpix1},
which is an equal-area pixelisation of the sphere, which we denote as $\Sspace^2$;
see Figure ~\ref{fig:healpix}.
Using the faces of the rhombic dodecahedron, we start by decomposing
the sphere into twelve patches.
Fine resolution is achieved by dividing these
patches into $N^2$ equal area pixels each, so that the total number
of pixels at this resolution is $12 \times N^2$.  At maximum resolution
$N = 2048$, the maps have about 50 million pixels.

\medskip\noindent\emph{Observed sky.}
The Planck satellite observes the sky at seven different frequency bands, leading to 
component-separated maps using four different techniques: Commander-Ruler (C-R), NILC, SEVEM, and SMICA; cf.\  \cite{planckCompSepMaps}.
We use these component-separated CMB maps throughout. These maps are contaminated by noise from various sources, including  inherent detector noise, and efforts   by the Planck team to denoise the data have not been completely successful.
Consequently,  our analysis is performed on the maps produced
by combining the CMB and  noise maps for each realization
of the simulation.
This is a fairly simple task, given that the map making exercise
is linear in nature:
\begin{align}
f_{final}  &=  f_{CMB} + f_{noise}.
\end{align}

\medskip\noindent\emph{Simulations.}
In addition to the observed data, the Planck team  released
a set of \emph{Full Focal Plane 8} (or FFP8) simulations \citep{planckSims}
of both  the CMB and noise. We use 1000 NILC simulations for our computational experiments.
These simulations  assume  that the CMB is a Gaussian random field,  consistent with the null-hypothesis of Gaussianity for the CMB, which is what we wish to check, and we  use them to estimate the error-bars for testing 
the significance of  differences between  observed and  simulated maps.

\medskip\noindent\emph{Degradation.}
In order to perform a scale-dependent analysis of the CMB maps,  we degrade them
to resolutions between $N = 1024$ and $8$, dividing $N$ by $2$ from
one level to the next.
The process of degradation amounts to decomposing them into spherical harmonics on the full sky at the input resolution. The spherical harmonics coefficients $a_{lm}$ are then convolved to the new resolution using the formula \citep{PlanckXXIII}
\begin{equation}
a_{lm}^{out} = \frac{b_l^{out}p_l^{out}}{b_l^{in}p_l^{in}}a_{lm}^{in},
\end{equation}
where $b_l$ is the beam transfer function, $p_l$ is the pixel window function, and \emph{in} and \emph{out} denote the input and the output functions at the different resolutions respectively. They are then synthesized into maps at the output resolution directly.

\medskip\noindent\emph{Masks.}
The observation of the CMB by the Planck satellite is incomplete
in some regions of the sky,  typically as a result of interference from  bright foreground objects, such as
our own  galactic disk and bright  point sources.
In these regions, the CMB sky map is reconstructed as a
constrained Gaussian field.
In order to avoid these areas in the analysis,
we use the most conservative UT78 mask released by the Planck team, see Figure~\ref{fig:UT78}.
It is a combination of all foreground objects with least sky coverage and so provides for a conservative analysis. The mask is a binary map, where reliable pixels of the CMB map are marked by the value 1, and the unreliable parts by 0.

For the scale dependent analysis, we also degrade the masks, so that the map and the mask have the same resolution. Degrading the original binary UT78 mask converts it into a non-binary mask in a thickened zone at the boundary separating the reliable part of the mask from the non reliable part.  Figure~\ref{fig:degradedNBMask} presents these yet to be binarized masks. To re-convert it to a binary mask, we set a range of binarization thresholds for our experiments: 0.7, 0.8, 0.9, 0.95. Pixels with values above or equal to the binarization threshold are marked as 1, and the rest as 0. 
Figure~\ref{fig:degradedBMask} presents the binarized maps at various degraded resolution, for binarization threshold 0.9. Table~\ref{tab:unmasked_area} presents the percentage of sky that is useable for analysis after masking at various resolutions, for this threshold. The percentage of usable area drops with decreasing resolution, with only $66\%$ for $N = 16$. Figure~\ref{fig:degradedMaskedMap} presents a visualization of the degraded and masked maps for all the resolutions analysed in this paper in the Mollweide projection view.

\begin{table} 
	\begin{center} 
		\begin{tabular}{|r|r|} 
			\hline
			Resolution   &   \% unmasked  \\ \hline \hline
			1024			&	77.19			\\ \hline 	
			512			&	76.52			\\ \hline 
			256			&	75.50			\\ \hline 
			128			&	73.37			\\ \hline 
			64			&	72.41			\\ \hline 
			32			&	69.39			\\ \hline 
			16			&	66.24			\\ \hline 
		\end{tabular} 
	\end{center} 
	\caption{Percentage of sky area covered by the unmasked regions for the various degraded resolutions between $N = 1024$ and $16$, for mask binarization threshold $0.9$.} 
	\label{tab:unmasked_area} 
\end{table} 

\begin{figure}
	\centering   
	\includegraphics[width=0.49\textwidth]{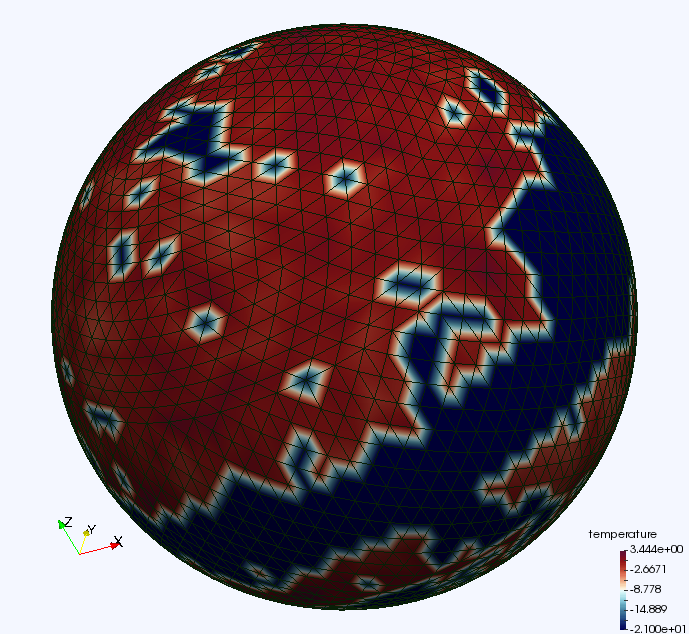}
	\caption{Visualization of the temperature field for the NILC observed maps at $N = 16$. Visible is also the corresponding triangulation, for which the pixel centers of the maps serve as the vertices. The temperature values are stored in the vertices of this triangulation.}
	\label{fig:triangulation}
\end{figure}

\subsection{Computational Pipeline}
\label{sec:computation}

The computational pipeline is tailored specifically to the Planck data.
The preprocessing step involves converting the CMB maps given in absolute units to a dimensionless unit, corrected for mean and scaled by the  standard deviation (computed using the non-masked pixels only). We use the \texttt{HealPix} package for the preprocessing step. The output of this operation is the normalized temperature values on $12\Res^2$ pixels, along with their coordinates on the sphere, which is the input to subsequent steps, which we discuss in five subsections:
(i) triangulating the surface of the sphere with the pixel centers as vertices,
(ii) sorting the vertices, edges, triangles to form an upper-star filteration,
(iii) computing the persistence in terms of a reduced boundary matrix,
(iv) computing the ranks of the relative homology groups $\relBetti{p}, p = 0, 1, 2$ and (v) the variationally optimal loops from the reduced matrix. The software is written in C++ and designed to handle very large data sets.

\medskip\noindent\emph{Triangulation.}
The \texttt{HealPix} format stores the data in twelve square
arrays of $\Res^2$ pixels each, with $\Res = 2048$ at the finest resolution.
The centers of these pixels are points on the faces of a rhombic dodecahedron.
With central projection, these points are mapped to the $2$-sphere and
distorted to achieve an approximately equal-area decomposition.
Taking the convex hull of these points in $\Rspace^3$, we get a
convex polytope whose boundary is homeomorphic to the sphere.
Most of the faces will be triangles, and the occasional faces with $k \geq 4$
sides can be subdivided into $k-2$ triangles to get a triangulation of the sphere. 
This triangulation is the input to all the down-stream computations.
Consisting of $V = 12 \Res^2$ vertices, $3V - 6$ edges, and $2V - 4$ triangles,
it represents the temperature field, $f \colon \Sspace^2 \to \Rspace$,
by storing the temperature value at every vertex.
We implicitly assume a piecewise linear interpolation
along the edges and the triangles.
Figure \ref{fig:triangulation} illustrates such a triangulation, using colours
to visualize the temperature field. We use the $\cgal$ library \citep{cgal} to implement the triangulation.

\medskip\noindent\emph{Upper-star filtration.}
Given a triangulation $K$ of $\Sspace^2$, let $K(\nu) \subseteq K$
contain all simplices (vertices, edges, triangles) whose temperature
values are $\nu$ or larger.
We use $K(\nu)$ as a proxy for $\Excursion (\nu)$, the corresponding excursion set.
Indeed, because of the linear interpolation, there is a deformation retraction from $\Excursion (\nu)$ to $K(\nu)$
\citep{EdHa10}, which implies that the two have corresponding components
and holes.
To process the sequence of excursion sets, it thus makes sense to sort
the vertices of $K$ in the order of decreasing temperature value.
More precisely, we order the simplices of $K$ such that
$\ssx$ precedes $\tsx$ if (i) $f(\ssx) > f(\tsx)$ or (ii) $f(\ssx) = f(\tsx)$ and $\dime{\ssx} < \dime{\tsx}$, in which $f(\ssx)$ is the minimum temperature value of the one, two, or
three vertices of $\ssx$.
The remaining ties are broken arbitrarily.
Assuming any two vertices have different temperature values,
then the edges and triangles that immediately follow a vertex
are exactly the ones in the \emph{upper star} of that vertex.
We therefore call any ordering that satisfies (i) and (ii) an
\emph{upper-star filter} of $K$ and $f$.
The corresponding \emph{upper-star filtration} consists of all
prefixes of the filter, each representing an excursion set.
This filtration is instrumental
in computing the persistence of components and holes.

\medskip\noindent\emph{Computing persistence.}
Given an upper-star filter of the piecewise linear temperature field,
there is optimized software available to compute its persistence \citep{PHAT}. We base our persistence computation on an adaptation of the software.
This software is a sophisticated implementation of the basic algorithm,
which we now describe and modify to get the variationally optimal
loops.
We write $\ssx_1, \ssx_2, \ldots, \ssx_n$ for the
simplices in the triangulation of the sphere,
sorted into an upper-star filter.
Let $\partial [1..n, 1..n]$ be the corresponding ordered
boundary matrix, with $\partial [i,j] = 1$, if $\ssx_i$ is a face
of $\ssx_j$ and $\dime{\ssx_i} = \dime{\ssx_j} - 1$,
and $\partial [i, j] = 0$, otherwise.
This matrix is sparse and stored as such.
The standard persistence algorithm reduces the matrix from left to right.
To reduce column $j$, we subtract columns to the left of $j$ with the goal
to move the lowest $1$ in column $j$ higher or eliminate it altogether.
We use modulo $2$ arithmetic, so subtracting is the same as adding:
$1-1 = 1+1 = 0$.
We call column $j$ \emph{reduced} if it is zero or its lowest $1$
has only $0$s in the same row to its left.
We modify the standard algorithm by continuing the reduction
even if the lowest $1$ cannot be changed any more,
calling the final result \emph{fully reduced}.
To be unambiguous, we explain this algorithm in pseudo-code,
where we write $\pivot{j}$ for the row index of the lowest $1$
in column $j$.

\begin{tabbing}
	mm\=m\=m\=m\=m\=m\=m\=m\=\kill
	\> \> {\tt for} $j=1$ {\tt to} $n$ {\tt do}                           \\*
	\> \> \> {\tt while} $\exists k<j$ with $\partial[\pivot{k}, j] = 1$ {\tt do} \\*
	\> \> \> \> add column $k$ to column $j$                              \\*
	\> \> \> {\tt endwhile}                                               \\*
	\> \> {\tt endfor}.
\end{tabbing}

\noindent The running time of this algorithm is cubic in the number of simplices
in the worst case, but the available optimized software is typically
much faster.

\medskip\noindent\emph{Ranks of relative homology groups.}
For computing the ranks of the homology groups relative to the mask, we set the vertices belonging to the mask at $+\infty$, and consider the complex $\Mask$, induced by the union of these vertices. This mask is closed by definition. We then compute the filtration and persistence diagram corresponding to absolute homology of $\Excursion \cup \Mask$. Writing $Dgm_p (\Excursion \cup \Mask)$ for the $p$-dimensional persistence diagram, and recalling that each diagram consists of intervals with real birth- and death-value, $b > d$, we get the ranks of homology groups relative to the mask:

\begin{align}
\relBetti{0}  &= \# \{[b,d) \in Dgm_0(\Excursion \cup \Mask) \mid  +\infty > b \geq \nu > d \} ; \\ \nonumber
\relBetti{1}  &= \# \{[b,d) \in Dgm_0(\Excursion \cup \Mask) \mid  +\infty = b > d \geq \nu \} \\ \nonumber
                  &+ \# \{[b,d) \in Dgm_1(\Excursion \cup \Mask) \mid  +\infty > b \geq \nu > d \} ;\\ \nonumber
\relBetti{2} &= \# \{[b,d) \in Dgm_1(\Excursion \cup \Mask) \mid  +\infty = b > d \geq \nu \} \\ \nonumber
                 &+ \# \{[b,d) \in Dgm_2(\Excursion \cup \Mask) \mid  +\infty > b \geq \nu > d \} .
\end{align}

\noindent For computing absolute homology, we set the mask pixels at $-\infty$ and consider the union of such vertices as the mask, which is open by definition.

\subsection{Statistical Tests}
\label{sec:stat}

The data consists of topological summaries ($\relBetti{0}, \relBetti{1}, \relEuler$) obtained from $1000$ simulations, as well as of the observed CMB field, processed according to NILC scheme. The goal is to estimate the probability that the physical model that produced the simulations would produce quantities consistent with those from the  observed CMB field. Let $\x_i \in \R^m$, $i=1,\ldots,n$, be a sample of i.i.d.\ $m$-dimensional  vectors,  drawn from a distribution $F$. Let $\y \in \R^m$ be another sample point, assumed to be drawn from a distribution $G$. We wish to test the (null) hypothesis that $F=G$, and shall give the test results in terms of \emph{$p$-values}, that compute the probability that $\y$ is `consistent' with this hypothesis. 
We consider two methods of testing for statistical consistency. The first is a parametric test based on the \emph{Mahalanobis distance}
\citep{mahalanobis}, also known as the \emph{$\chi^2$-test}.
The second is a non-parametric test based on the \emph{Tukey depth} \citep{depth}.
The $\chi^2$-test is more standard but has the disadvantage of assuming
that the compared quantities follow a Gaussian distribution,
while the Tukey depth works without any assumption on the distribution.

\medskip\noindent\emph{Mahalanobis distance or $\chi^2$-test.}
Let $\bar{\x} = \sum_{i=1}^{n} \x_i/n$ and $\S = \sum_{i=1}^{n} (\x_i - \bar{\x}) (\x_i - \bar{\x})^\T / (n-1)$ the sample mean and covariance matrix of the sample $\x_1,\ldots,\x_n$. The squared Mahalanobis distance of $\y$ to  $\bar{\x}$ is then

\begin{equation}
d^2_{\rm Mahal}(\y) = (\y - \bar{\x})^\T \S^{-1} (\y - \bar{\x}).
\label{eq:Mahal}
\end{equation}

If $F$ is assumed to be Gaussian and $n$ is large,  then under the hypothesis that $G=F$ the squared Mahalanobis distance \eqref{eq:Mahal} is approximately distributed as a   $\chi^2$ distribution with $m$ degrees of freedom. Thus the corresponding  $p$-value is

\begin{equation}\label{eq:Mahal-p}
p_{\rm Mahal}(\y) = P[\chi^2_m > d^2_{\rm Mahal}(\y)].
\end{equation}

\medskip\noindent\emph{Tukey depth.}
As will be shown in the data analysis, the distribution $F$ does not always conform to elliptical contours and therefore may not be Gaussian. In such a setting,  $p$-values computed using the Mahalanobis distance may not be reliable.

The Tukey half-space depth provides a general metric  for identifying outliers  in a flexible manner and in a non-parametric setting. Take $\x_i$, $i=1,\dots,n$ and  $\y$ as above, making no assumptions on the structure of $F$ and $G$, and let $\mathbf{z}$ be any point in $\mathbb R^m$.  Then the half-space depth $d_{\rm dep}(\mathbf{z}; \x_1,\ldots,\x_n)$ of $\mathbf{z}$ within the sample of the $\x_i$ is the smallest fraction of the $n$ points $\x_1,\ldots,\x_n$ to either side of any hyperplane passing through $\mathbf{z}$. By definition, the half-space depth is a number between 0 and 0.5. Points that have the same depth constitute a non-parametric estimate of the  isolevel contour of the distribution $F$. 

To evaluate a $p$-value for $\y$, we first compute $d_j = d_{\rm dep}(\x_j; \x_1,\ldots,\x_n)$ for every point $\x_j$, $j=1,\ldots,n$, yielding an empirical distribution of depth. Then the $p$-value is computed as the proportion of points whose depth is lower than that of $\y$:

\begin{equation}\label{eq:hsd-p}
p_{\rm dep} (\y)  =  \#\{j \mid d_j > d_{dep}(y) \} / n
\end{equation}

Note that, by construction, the depth $p$-value increases in units of $1/n$. For computing half-space depths  below, we use the {\tt R} package {\tt depth}.

\section{Results}
\label{sec:3}

We use the Planck maps for our analyses, which
measure fluctuations about the mean in the CMB temperature to an accuracy
of $10^{-5}$K \citep{planckMain}. 
We compare observed maps with $1000$ Full Focal Plane 8 (FFP8) simulations cleaned using the NILC technique \citep{planckSims}. The simulations are based on the LCDM paradigm and assume that the temperature fluctuations have a Gaussian distribution. In addition, we also compare the observed maps cleaned using the C-R, SEVEM and SMICA techniques with the NILC simulations, and note that they exhibit similar characteristics. However, since these maps arise from different processing techniques, the results, while broadly reproducing and corroborating those based on the NILC technique, are subject to various interpretations, and hence we abstain from discussing them. We perform our analyses for a range of scales between $0.05$ and $7.33$ degrees, which correspond to resolutions between $\Res = 1024$ and $\Res = 8$ in the \texttt{HealPix} format \citep{Healpix1}. Further degradation of the maps destabilizes the statistics due to the low number of data points in these cases. We do this for a range of mask binarization thresholds: 0.7, 0.8, 0.9 and 0.95. See Section~\ref{sec:data} for the details of degradation and masking. 

We present our analyses in terms of the ranks of relative homology groups, $\relBetti{p}$ for $0 \leq p \leq 1$. The relative components and loops are quantified by the \emph{relative component function}, $\relBetti{0}: \Rspace \to \Rspace$, and the \emph{relative loop function}, $\relBetti{1}: \Rspace \to \Rspace$.
We present the graphs of $\relBetti{0}$, $\relBetti{1}$, and of the (relative) Euler characteristic, $\relEuler$, followed by statistical tests that estimate the significance of results. If $f(x): \Sspace^2 \to \Rspace$ is the absolute temperature at a location $x$, and $f_0$ the mean temperature of the distribution, the dimensionless temperature is given by: $\nu (x)  =  (f(x) - f_0)/\sigma (f)$,
where $\sigma (f)$ is the standard deviation computed from the non-masked pixels. We then obtain the ranks of relative homology groups as functions of the normalized temperature. 

\begin{figure*}
	\centering 
	\rotatebox{-90}{\includegraphics[height=0.9\textwidth]{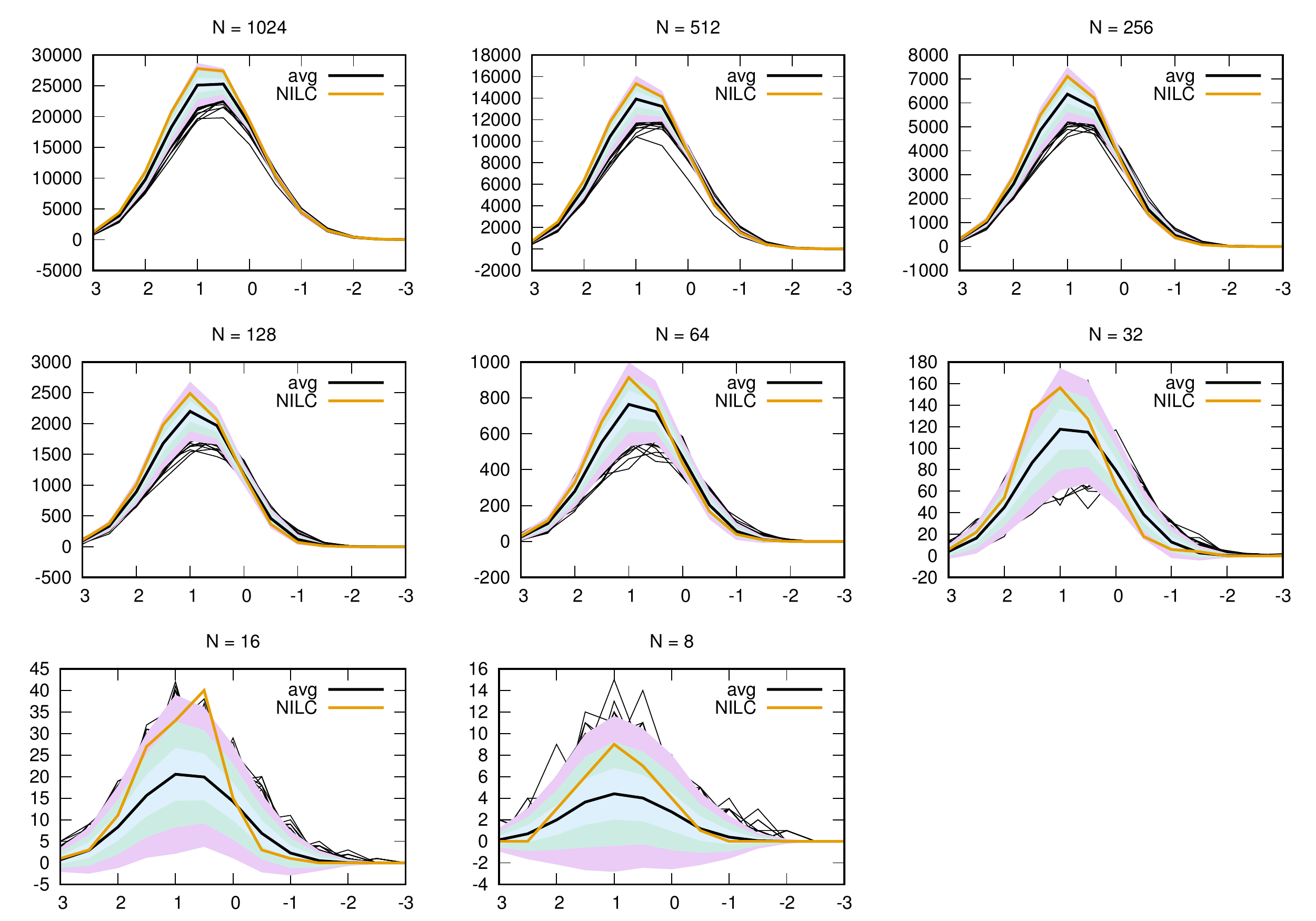}}\\
	\rotatebox{-90}{\includegraphics[height=0.9\textwidth]{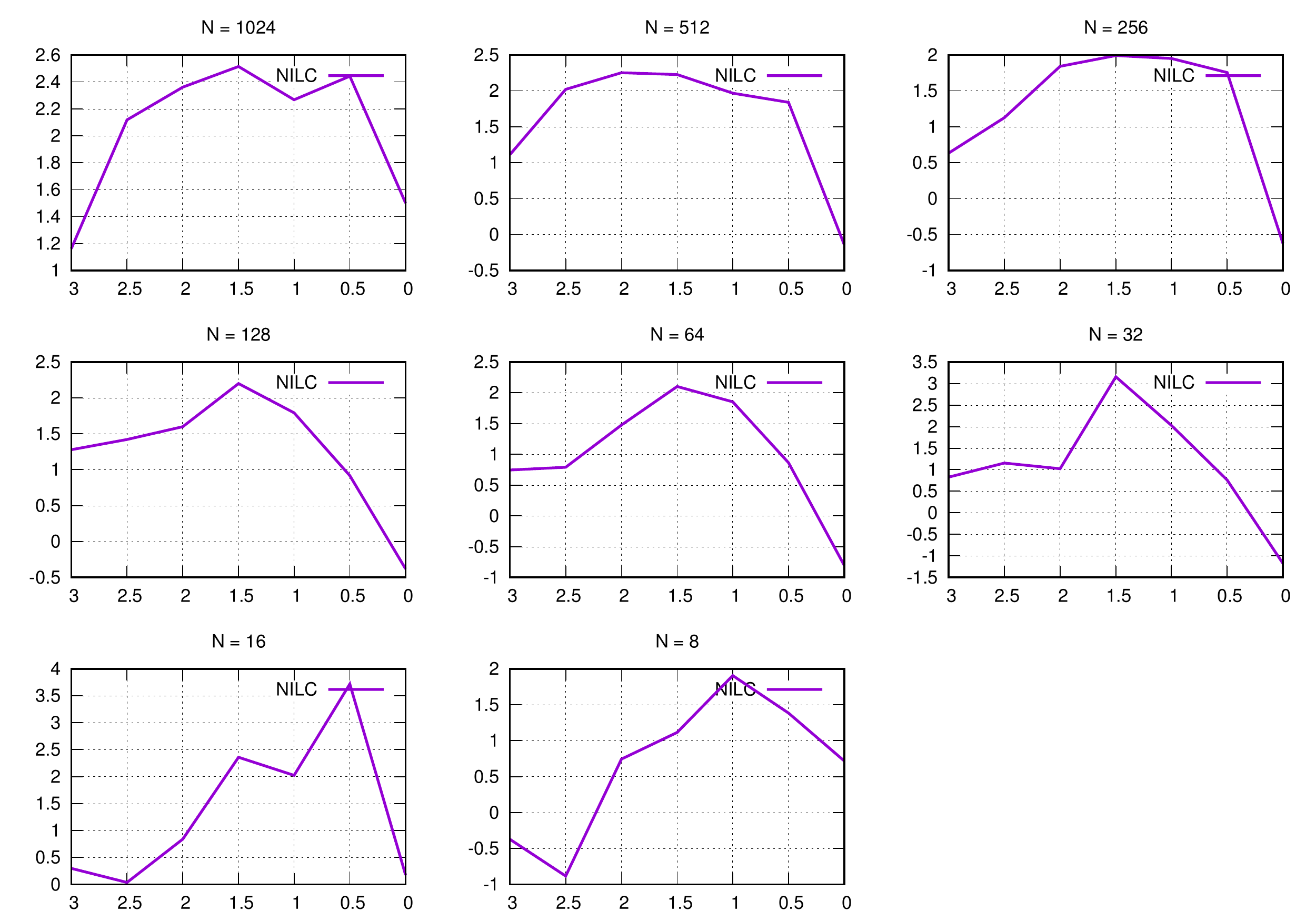}}\\
	\caption{Figure presenting the $\relBetti{0}$ graphs for resolutions between $N = 1024$ and $N = 8$. Top three rows: The observed (yellow) curve, and the expected (black) curve computed from 1000 simulations, along with bands drawn up to 3$\sigma$. Also plotted underneath are the 1000 curves from individual simulations. Bottom three rows: curve presenting the significance of difference between observations and simulations for the various temperature thresholds. Maximum noted deviation is at $N = 16$ at $3.7\sigma$. The threshold along the horizontal axis runs from positive to negative, in view of the fact that we analyse superlevel sets of the normalized temperature field.}
	\label{fig:b0_graph}
\end{figure*}

\begin{figure*}
	\centering 
	\rotatebox{-90}{\includegraphics[height=0.9\textwidth]{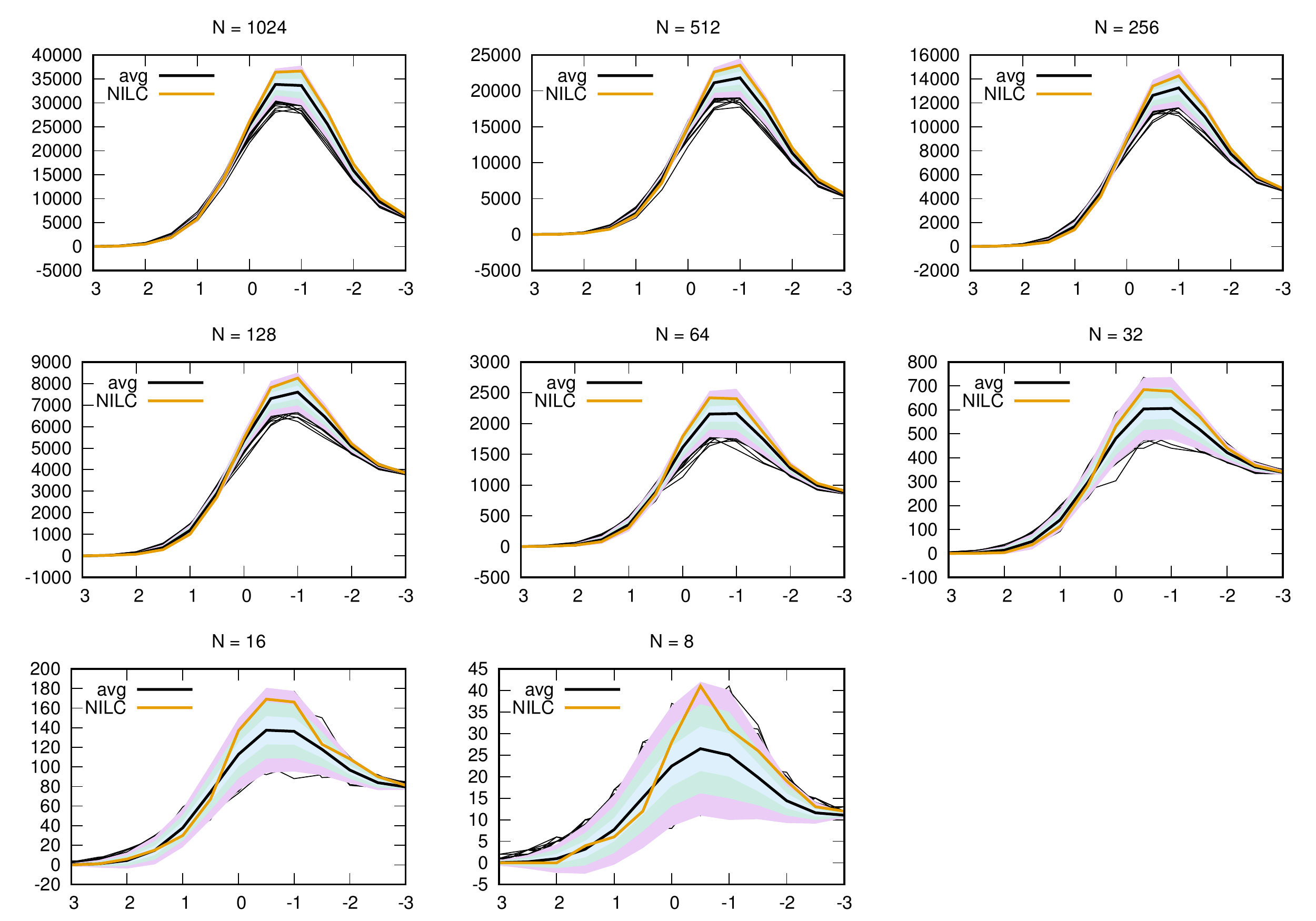}}\\
	\rotatebox{-90}{\includegraphics[height=0.9\textwidth]{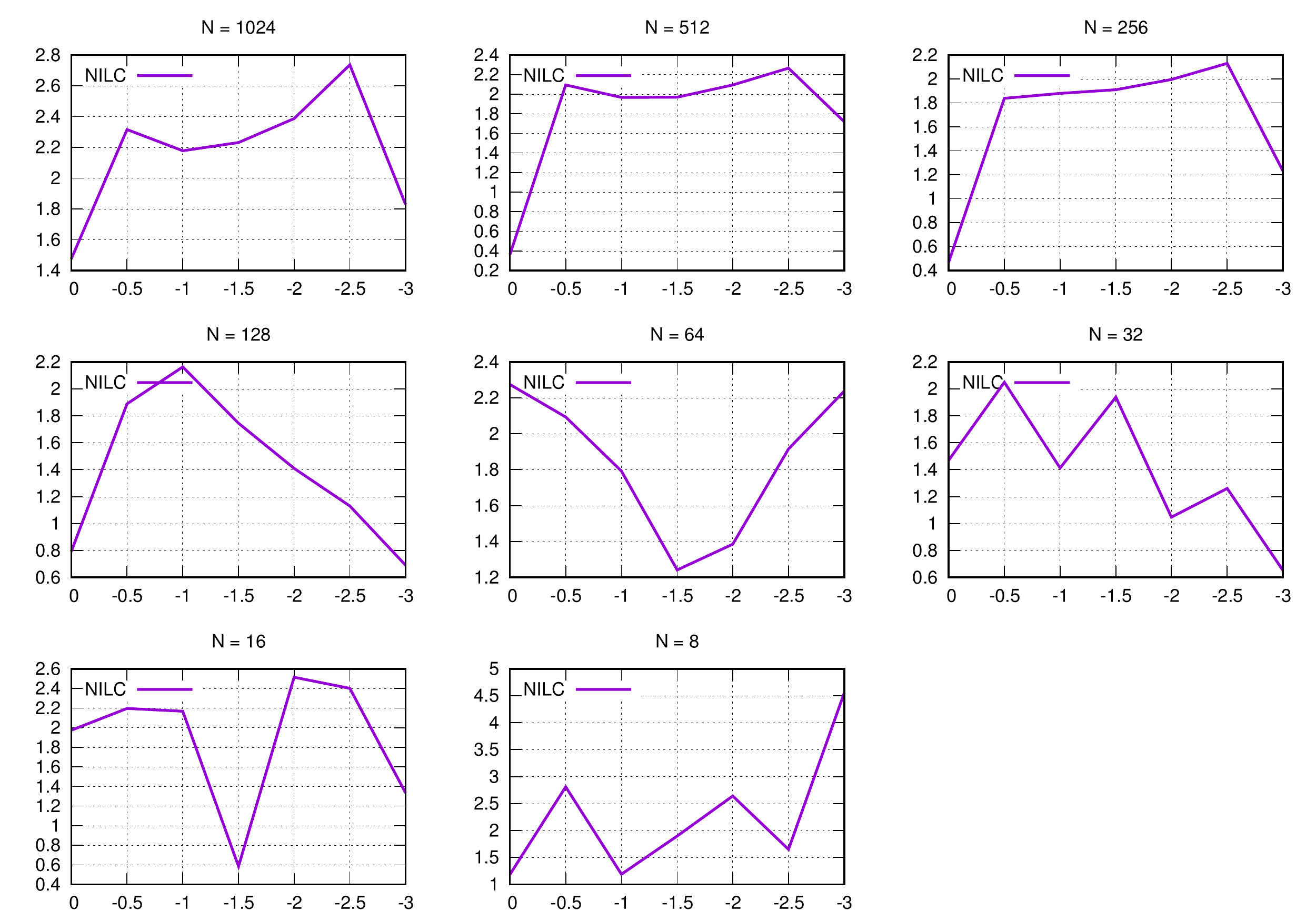}}\\
	\caption{Figure presenting the $\relBetti{1}$ graphs for resolutions between $N = 1024$ and $N = 8$. Top three rows: The observed (yellow) curve, and the expected (black) curve computed from 1000 simulations, along with bands drawn up to 3$\sigma$. Also plotted underneath are the 1000 curves from individual simulations. Bottom three rows: curve presenting the significance of difference between observations and simulations for the various temperature thresholds. Maximum noted deviation is at $N = 8$ at $2.9\sigma$. The threshold along the horizontal axis runs from positive to negative, in view of the fact that we analyse superlevel sets of the normalized temperature field.}
	\label{fig:b1_graph}
\end{figure*}

\begin{figure*}
	\centering 
	\rotatebox{-90}{\includegraphics[height=0.9\textwidth]{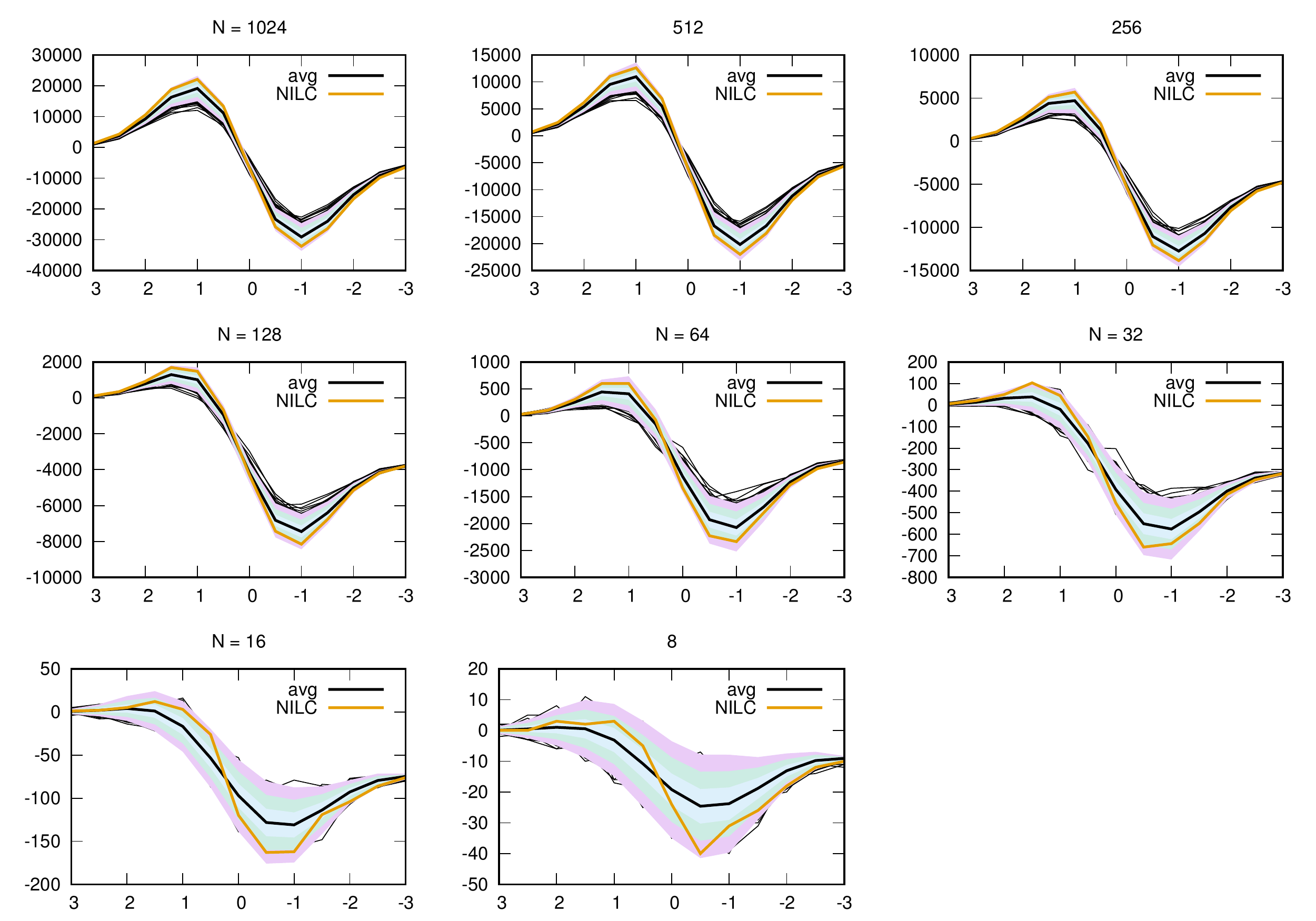}}\\
	\rotatebox{-90}{\includegraphics[height=0.9\textwidth]{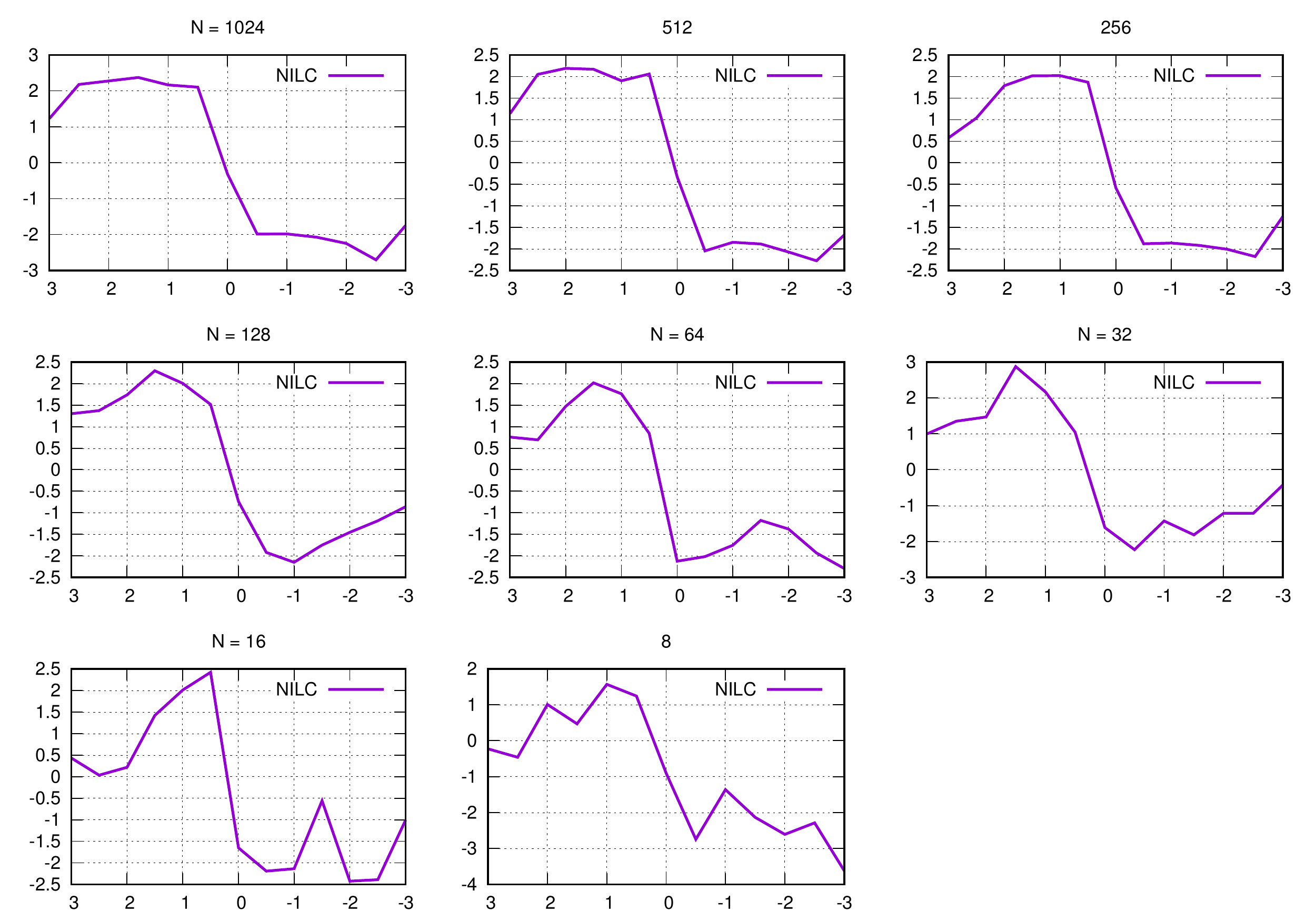}}\\
	\caption{Figure presenting the Euler characteristic graphs for resolutions between $N = 1024$ and $N = 8$. Top three rows: The observed (yellow) curve, and the expected (black) curve computed from 1000 simulations, along with bands drawn up to 3$\sigma$. Also plotted underneath are the 1000 curves from individual simulations. Bottom three rows: curve presenting the significance of difference between observations and simulations for the various temperature thresholds. Maximum noted deviation is at $N = 32$ at $2.8\sigma$. The threshold along the horizontal axis runs from positive to negative, in view of the fact that we analyse superlevel sets of the normalized temperature field.}
	\label{fig:EC_graph}
\end{figure*}

\subsection{Ranks of relative homology groups}

To carry out omnibus tests, we choose 13 a-priori levels, $\ell_{-6},\dots,\ell_6$, where 
$\ell_k = k/2$, so that the normalized temperature thresholds run from $-3$ to $+3$ in steps of $0.5$, and consider collections of random variables 
$\relBetti{0}(\ell_k)$, $\relBetti{1}(\ell_{k})$, and $\relEuler(\ell_{k})$, for $-6 \le k \le 6$.

Figures~\ref{fig:b0_graph}--\ref{fig:EC_graph} show the curves of $\relBetti{0}$, $\relBetti{1}$, and $\relEuler$ for resolutions between $\Res = 1024$ and $\Res = 8$, for mask threshold 0.9. The graphs present the average curve (black) from $1000$ NILC simulations, with error-bands drawn up to $3\sigma$. The individual curves from simulations are drawn in dotted black, a few of which escape the $3\sigma$ band. Also plotted are curves from NILC observed maps in yellow.  $\relBetti{0} (\nu)$, $\relBetti{1} (\nu)$ and $\relEuler (\nu)$ show a difference from simulations peaking near $2\sigma$ for some temperature levels for all resolutions. Additionally, $\relBetti{0}$ and $\relBetti{1}$  show differences peaking between  $3-4\sigma$ sporadically between $N = 32$ and $N = 8$. 

\subsection{Experimental evidence of Euler characteristic suppression}

As noted before, the Euler-Poincar\'{e} formula states that the Euler characteristic is the alternating sum of the Betti numbers. As a consequence, the signals in Euler characteristic are suppressed by design, due to the cancellation of the constituent Betti numbers. Our experiments provide evidence for such suppressions of the topological signals emanating from the Euler characteristic. As an example, consider the quantities at the degraded resolution $N = 16$, and temperature threshold value $\nu = 0.5$. \relBetti{0} at this resolution and threshold has a significant difference between observations and simulations at $3.7\sigma$ (Figure~\ref{fig:b0_graph}), but the corresponding value for the Euler characteristic is $2.4\sigma$ (Figure~\ref{fig:EC_graph}). This is because of the cancellation effects between $\relBetti{0}$ and $\relBetti{1}$ in determining the Euler characteristic. One may find more instances of such cancellation effects in the graphs.

\subsection{Statistical significance of the results}

\begin{table}
	\tiny \centering
	\begin{tabular}{|r||rrr|rrr|} \hline
		
		\multicolumn{7}{|c|}{Relative homology} \\
		\hline
		
		& \multicolumn{3}{|c|}{Mahalanobis}  & \multicolumn{3}{|c|}{Tukey Depth} \\
		
		resolution & \multicolumn{1}{c}{$\relBetti{0}$} & \multicolumn{1}{c}{$\relBetti{1}$}
		
		& \multicolumn{1}{c}{$\relEuler$}
		
		& \multicolumn{1}{|c}{$\relBetti{0}$} & \multicolumn{1}{c}{$\relBetti{1}$}
		
		& \multicolumn{1}{c|}{$\relEuler$} \\ \hline \hline
		
		\multicolumn{7}{|c|}{threshold = 0.70} \\ \hline
		
		1024       & 0.236	& 0.244 &0.472 & $\bf{<0.001}$ & $\bf{<0.001}$ & 0.302 \\
		
		512       & 0.492	& 0.325&0.666 & 0.134 & $\bf{<0.001}$ & 0.685 \\
		
		256       & 0.602	& 0.513 & 0.760 & 0.201 & 0.211 &0.579 \\
		
		128       & 0.260	& 0.441 & 0.627 & $\bf{<0.001}$ & 0.171 &0.327 \\
		
		64       & 0.335	& 0.278& 0.528 & 0.171 & $\bf{<0.001}$ &$\bf{<0.001}$ \\
		
		32       & 0.082	&0.302 &0.442 & $\bf{<0.001}$ & $\bf{<0.001}$&$\bf{<0.001}$ \\
		
		16       & \bf{0.018} & \bf{0.030} &0.120 & $\bf{<0.001}$ & $\bf{<0.001}$ &$\bf{<0.001}$  \\
		
		8       & 0.373 & $\bf{<0.001}$ &\bf{0.012} & 0.430 & $\bf{<0.001}$ &$\bf{<0.001}$ \\ \hline
		
		summary & \bf{0.002}& \bf{0.001}& \bf{0.002}& $\bf{<0.001}$& $\bf{<0.001}$& $\bf{<0.001}$  \\ \hline\hline
		
		\multicolumn{7}{|c|}{threshold = 0.80} \\ \hline
		
		1024 	& 0.225 & 0.278 &0.472 &$\bf{<0.001}$ & $\bf{<0.001}$ &0.410 	\\
		
		512   	& 0.499  & 0.348 &0.686 & 0.130& $\bf{<0.001}$ &0.649 \\
		
		256  	& 0.559  & 0.481 &0.679 & 0.139 &0.218 &0.538 \\
		
		128   	& 0.295 & 0.484 &0.633 & $\bf{<0.001}$ &$\bf{<0.001}$ &0.331 \\
		
		64 	& 0.250 & 0.269 &0.382 & $\bf{<0.001}$ &$\bf{<0.001}$ &$\bf{<0.001}$ \\
		
		32 	& 0.082 & 0.406 &0.538 & $\bf{<0.001}$ &$\bf{<0.001}$ &$\bf{<0.001}$\\
		
		16    	& \bf{0.024} & \bf{0.043} &0.082 & $\bf{<0.001}$ &$\bf{<0.001}$ &$\bf{<0.001}$ \\
		
		8    	& 0.202 & $\bf{<0.001}$ &\bf{0.013} & 0.142 &$\bf{<0.001}$ &0.220\\ \hline
		
		summary	& \bf{0.001}& \bf{0.001}&\bf{0.002}& $\bf{<0.001}$& \bf{0.032}&$\bf{<0.001}$ \\ \hline\hline
		
		\multicolumn{7}{|c|}{threshold = 0.90} \\ \hline
		
		1024 	& 0.225& 0.278 &0.472 &$\bf{<0.001}$& $\bf{<0.001}$&0.410 	\\
		
		512   	& 0.526 &0.340 &0.661 & 0.264 &$\bf{<0.001}$ &0.641 \\
		
		256  	& 0.601 &0.584 &0.738 & 0.150 &0.306 &0.589 \\
		
		128   	& 0.259 &0.525 &0.486 & $\bf{<0.001}$ &0.160 &$\bf{<0.001}$ \\
		
		64 	& 0.611 &0.248 &0.444 & 0.547 &$\bf{<0.001}$ &0.428 \\
		
		32 	& 0.053 &0.305 &0.300 &$\bf{<0.001}$ &0.243 &0.340 \\
		
		16    	& \bf 0.009 & 0.054 &0.081& $\bf{<0.001}$ &$\bf{<0.001}$ &$\bf{<0.001}$ 	\\
		
		8    	& 0.408 & $\bf{<0.001}$ &\bf 0.007 & 0.401 &$\bf{<0.001}$&$\bf{<0.001}$\\ \hline
		
		summary	& \bf{0.010}& $\bf{<0.001}$& \bf{0.001}& $\bf{<0.001}$& \bf{0.002}& $\bf{<0.001}$\\ \hline\hline

		\multicolumn{7}{|c|}{threshold = 0.95} \\ \hline
		
		1024 	& 0.225& 0.278 &0.472 & $\bf{<0.001}$ &$\bf{<0.001}$ &0.410 	\\
		
		512   	& 0.531 &0.340 &0.654 & 0.207 &$\bf{<0.001}$ &0.445\\
		
		256  	& 0.602 &0.550 &0.702 & 0.139 &0.345 &0.434 \\
		
		128   	& 0.309 &0.524 &0.554& $\bf{<0.001}$ &0.165 &0.446 \\
		
		64 	& 0.420 &0.157 &0.430 & 0.276 &$\bf{<0.001}$ &$\bf{<0.001}$ 	\\
		
		32 	& 0.076 &0.246 &0.174& $\bf{<0.001}$ &$\bf{<0.001}$ &$\bf{<0.001}$\\
		
		16    	& 0.436 &\bf0.016 &\bf0.027 & 0.383 &$\bf{<0.001}$&$\bf{<0.001}$\\
		
		8    	& 0.871 &\bf0.047 &0.188 & 0.846 &$\bf{<0.001}$ &0.517\\ \hline
		
		summary	& 0.213& \bf{0.003}&\bf{0.010}& $\bf{<0.001}$& \bf{0.009}&\bf{0.001}	\\ \hline

	\end{tabular}
	\caption{Table displaying the two-tailed $p$-values for relative homology obtained from parametric (Mahalanobis distance) and non-parametric (Tukey depth) tests, for four mask binarization thresholds. The last entry for each threshold is the summary statistic computed across all resolutions. Marked in boldface are $p$-values $0.05$ or smaller.} 
	\label{tab:degrade-pvalues_1}
	
\end{table}

We consider the two methods detailed in Section~\ref{sec:stat}, and
present $p$-values of the observed maps for both.
We consider the variables $\relBetti{0}(\ell_{k=0,\ldots,6})$, $\relBetti{1}(\ell_{k=-6,\ldots,0})$, and $\relEuler(\ell_{k=-6,\ldots,6})$ for estimating the statistical significance of the results.
The choice of regions is determined by the fact that $\relBetti{0}(\nu)$  tends to be  small, and carries little information for $\nu<0$, $\relBetti{1}(\nu)$ tends to be small for $\nu>0$, while the Euler characteristic is informative over the full range of levels.
We perform summary and specific tests for mask binarization threshold values corresponding to 0.7, 0.8, 0.9 and 0.95. For the summary tests, we take the full set of quantities for all the resolutions between $N = 1024$ and $N = 8$, and the relevant thresholds, as vectors separately for each of the three topological quantities. The last entry for each mask threshold in Table~\ref{tab:degrade-pvalues_1} presents the summary $\chi^2$ and depth $p$-values. Overall, there is ample indication that the observations differ from the simulations. This is followed by specific tests for each resolution. The rest of the entries in Table~\ref{tab:degrade-pvalues_1} present the Mahalanobis and the depth $p$-values for each resolution, for different mask thresholds. The Mahalanobis distances are particularly small for $\relBetti{1}$ at $N = 16$, and very significant at $N = 8$, across all binarization thresholds. $\relBetti{0}$ and $\relEuler$ also show significance, but are not stable across binarization thresholds. The depth $p$-values are very significant for $\relBetti{1}$ for $N = 16$ and $N = 8$, while $\relBetti{0}$  shows high significance at $N = 32$ consistently across the range of binarization thresholds. The depth $p$-values also show high significance at higher resolutions, more often for $\relBetti{1}$ than for \relBetti{0}, but not at all for $\relEuler$, presumably because of cancellation effects. $\relBetti{1}$ shows significance more often than $\relBetti{0}$ and $\relEuler$ when considering Tukey depth, and is an order of magnitude more significant compared to $\relBetti{0}$ and $\relEuler$, when considering the Mahalanobis distance. 

\begin{figure*}
	\centering   
	\includegraphics[width=\textwidth]{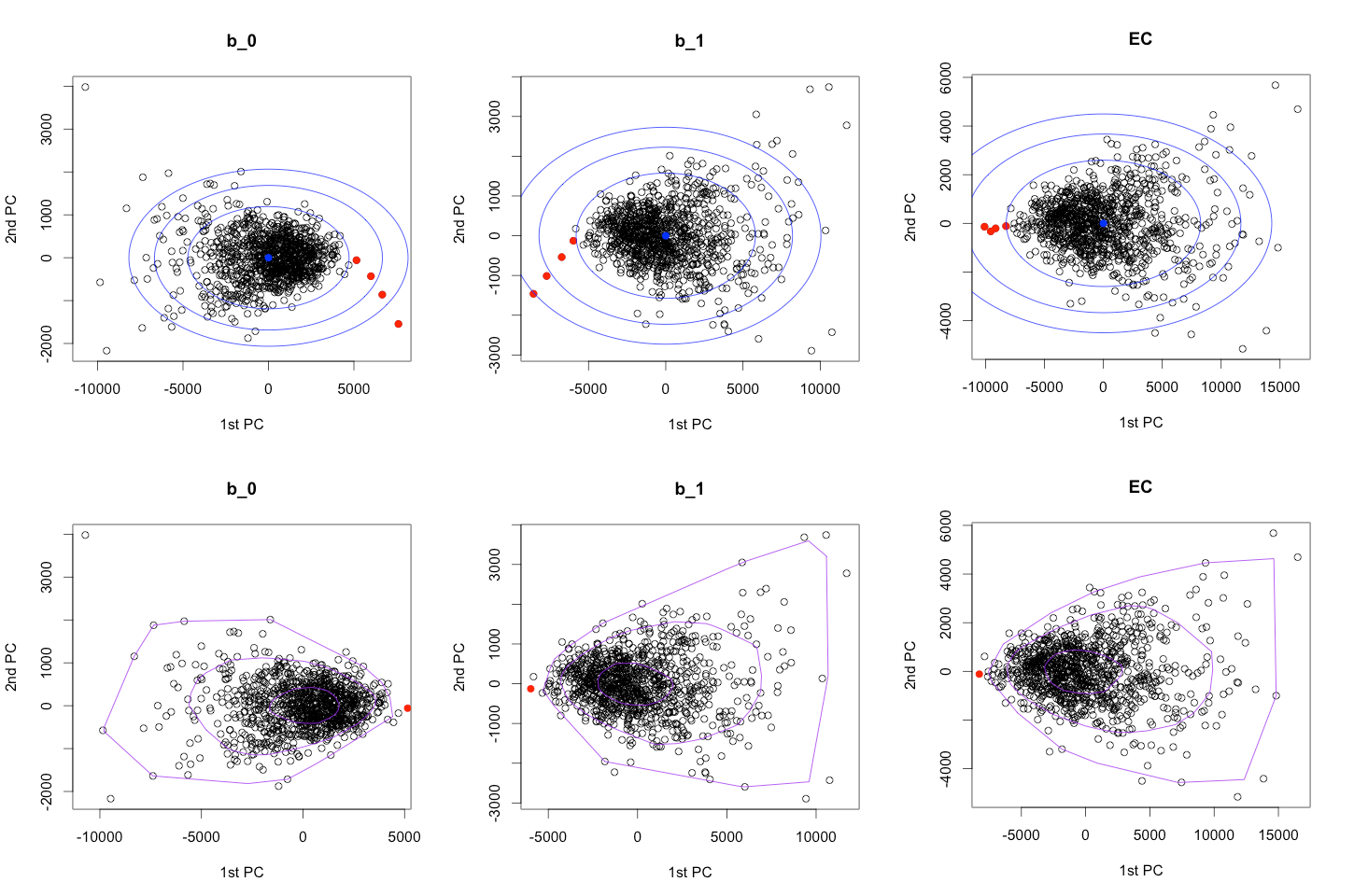}
	\caption{Summary test. Projection onto first two principal components. Mahalanobis and depth contours corresponding to $p$-values of 0.1, 0.01 and 0.001 are shown in blue (top) and purple (bottom). Observed CMB points are in red.}
	\label{fig:globalContours}
\end{figure*}

\begin{figure*}
	\centering  
	\includegraphics[width=\textwidth]{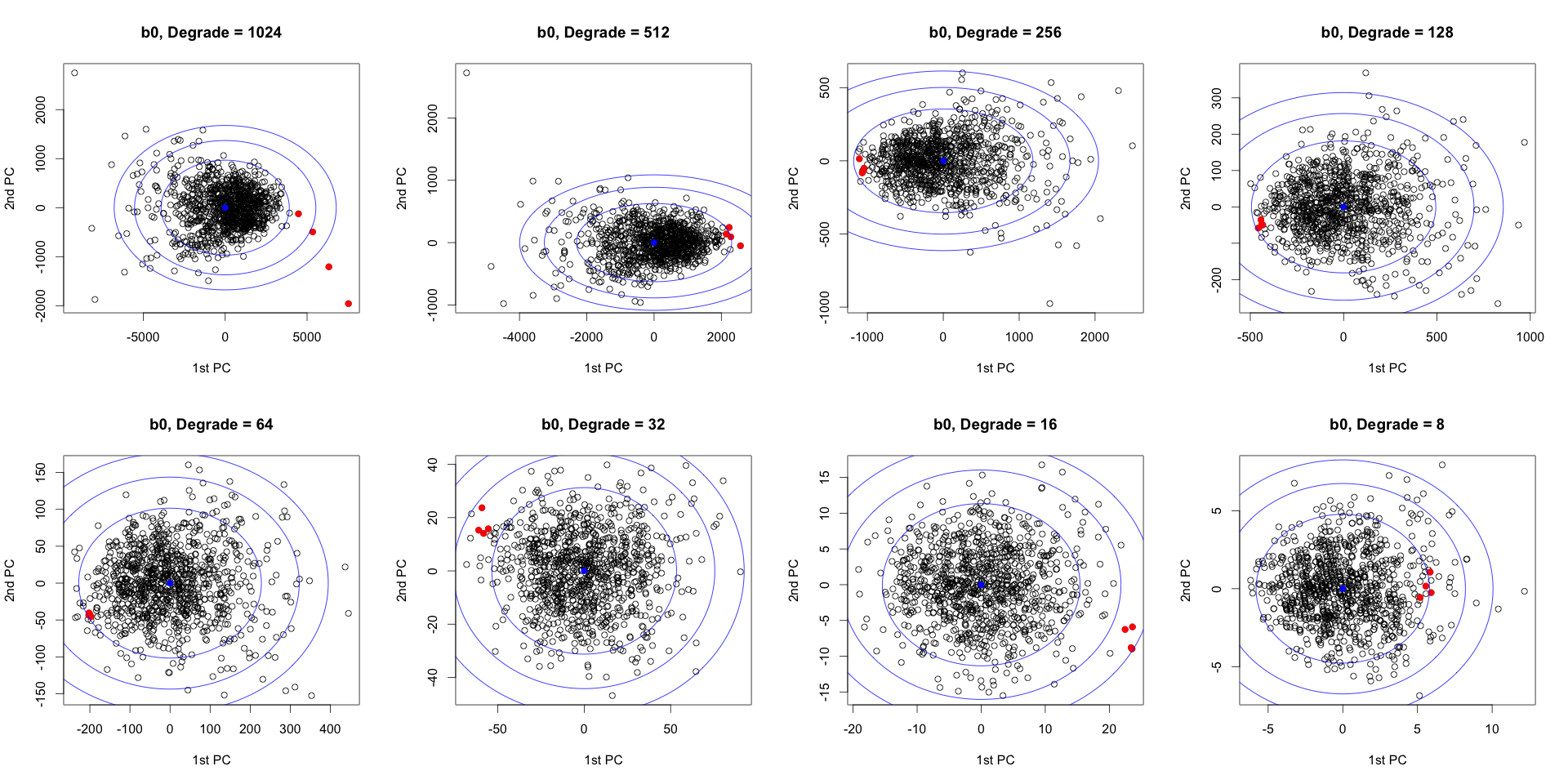}\\
	\includegraphics[width=\textwidth]{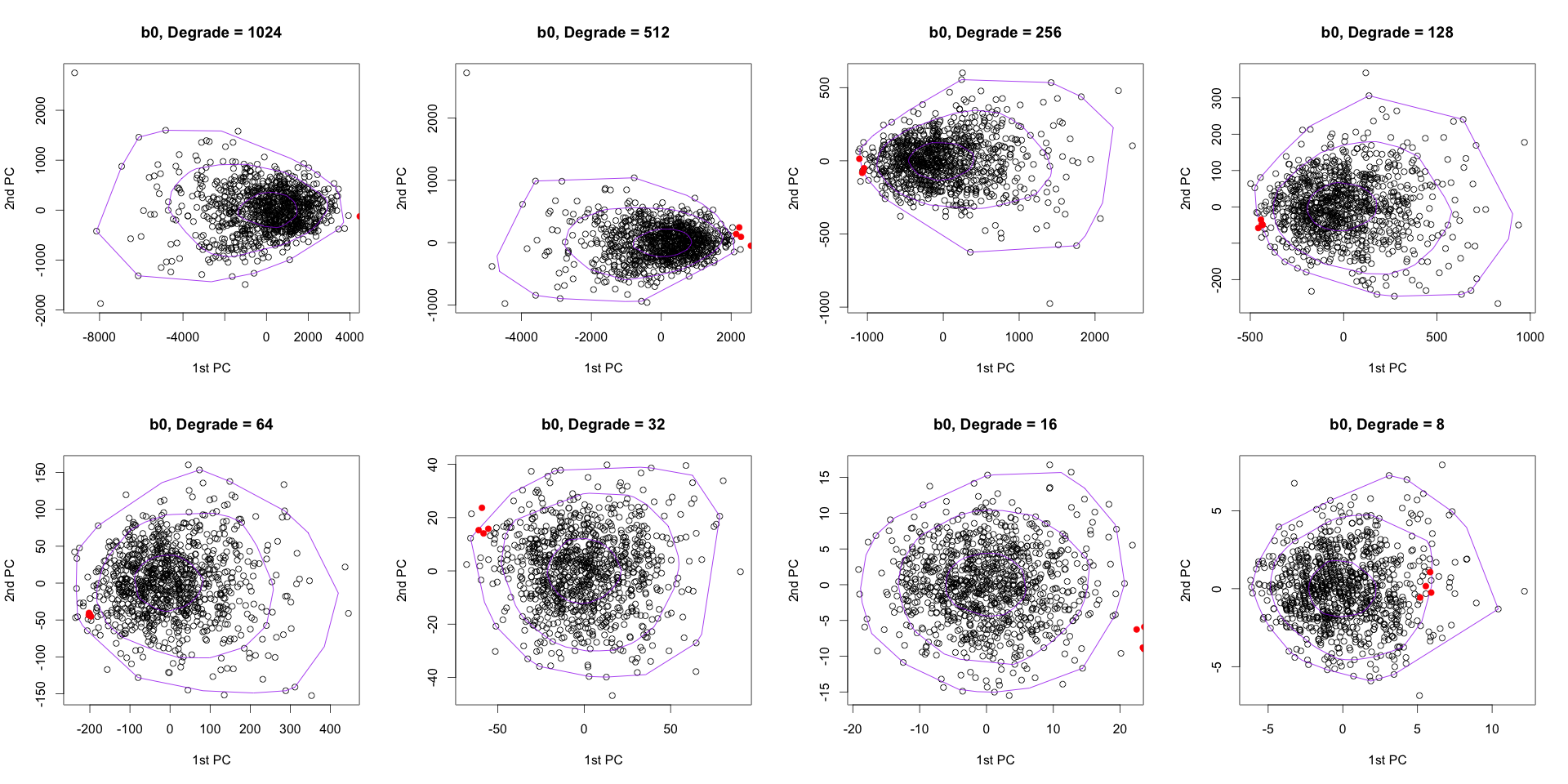}\\
	\caption{ Projection onto first two principal components for specific resolution tests for $\relBetti{0}$, $\relBetti{1}$, $\relEuler$.  Mahalanobis and depth contours corresponding to $p$-values of 0.1, 0.01 and 0.001 are shown in blue (top three rows) and purple (bottom three rows). Observed CMB points are in red.}
	\label{fig:specificContours_b0}
\end{figure*}

\begin{figure*}
	\centering  
	\includegraphics[width=\textwidth]{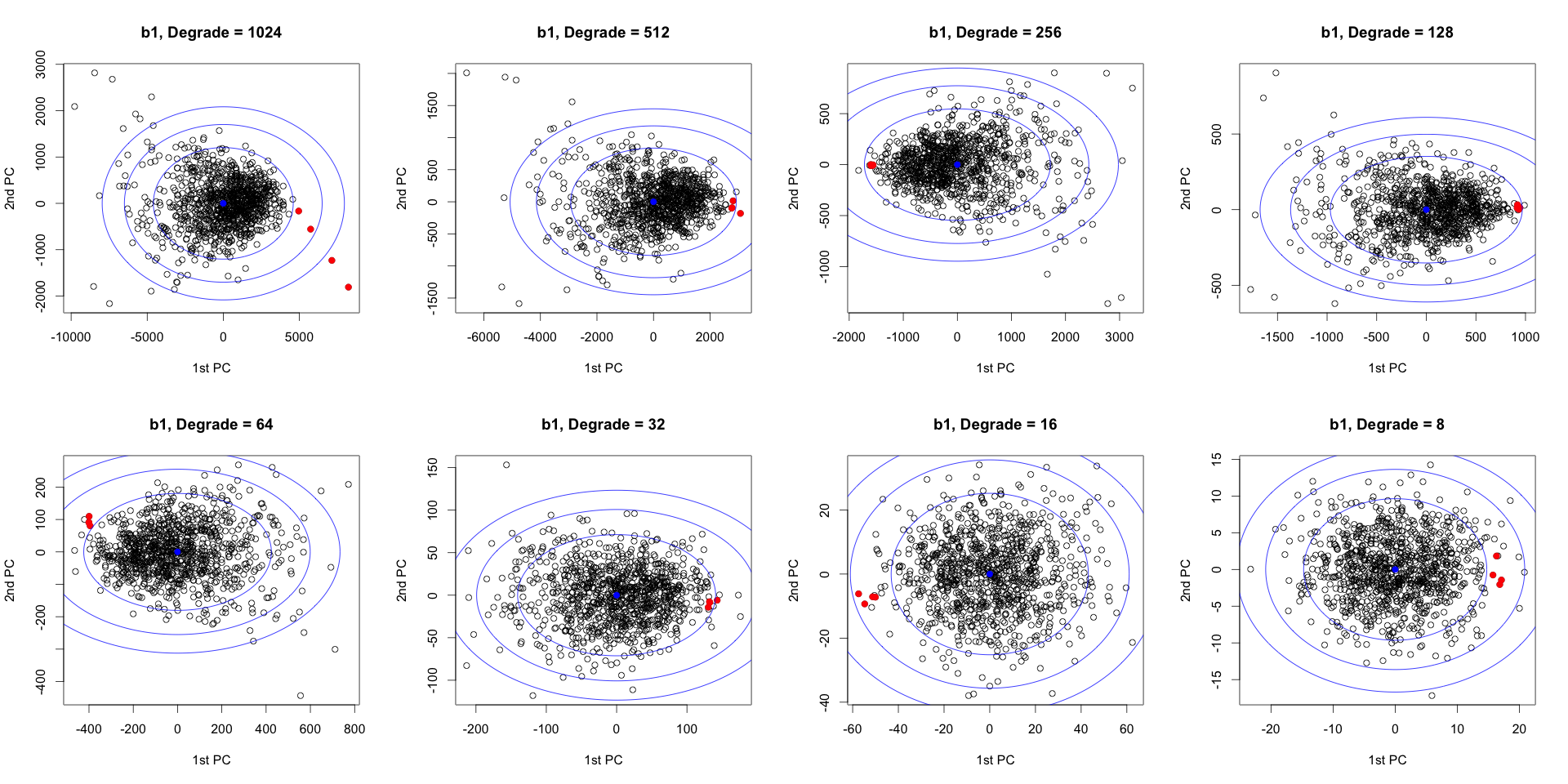}\\
	\includegraphics[width=\textwidth]{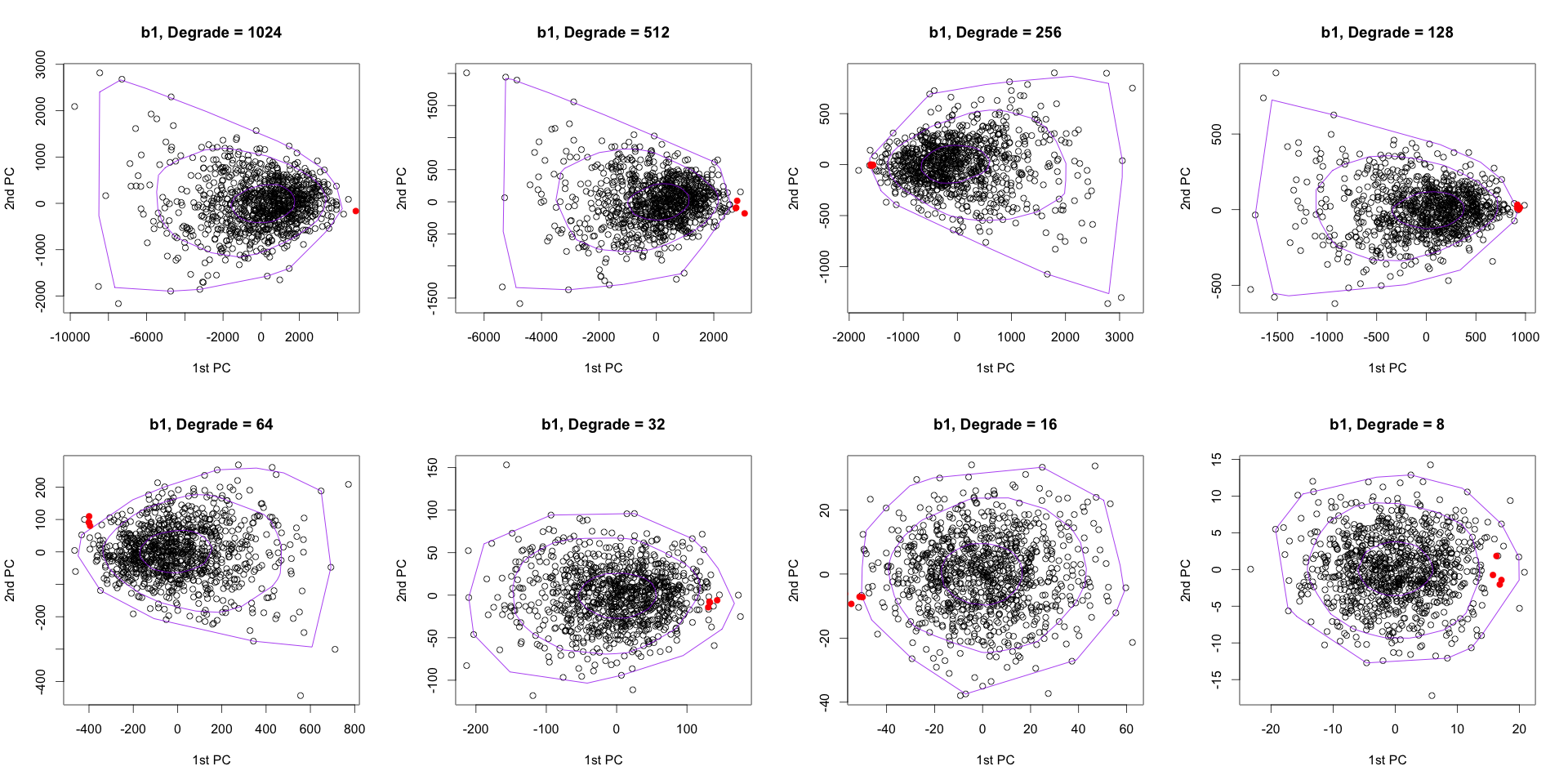}\\
	\caption{ Projection onto first two principal components for specific resolution tests for  $\relBetti{1}$.  Mahalanobis and depth contours corresponding to $p$-values of 0.1, 0.01 and 0.001 are shown in blue (top three rows) and purple (bottom three rows). Observed CMB points are in red.}
	\label{fig:specificContours_b1}
\end{figure*}

\begin{figure*}
	\centering  
	\includegraphics[width=\textwidth]{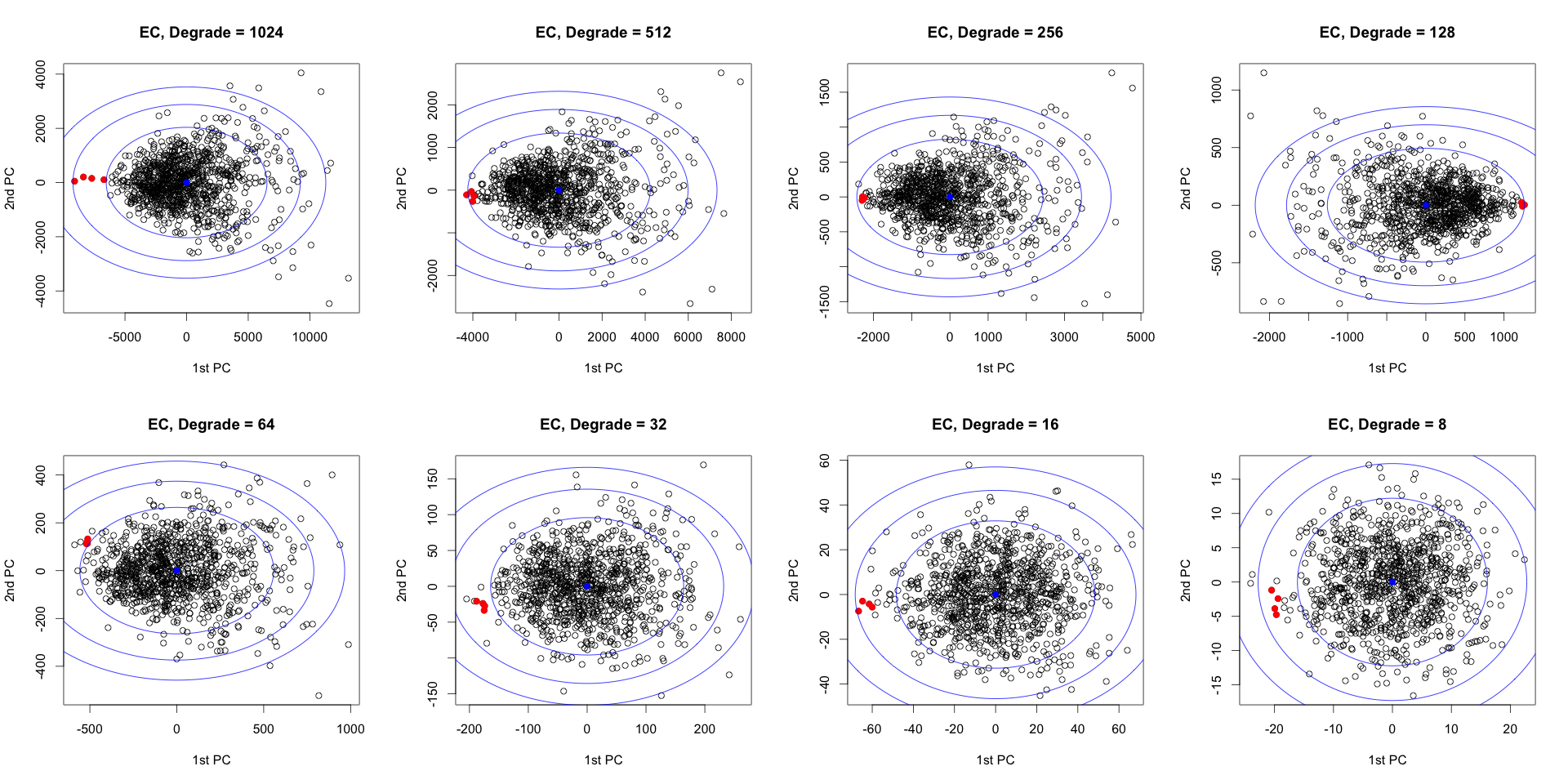}\\
	\includegraphics[width=\textwidth]{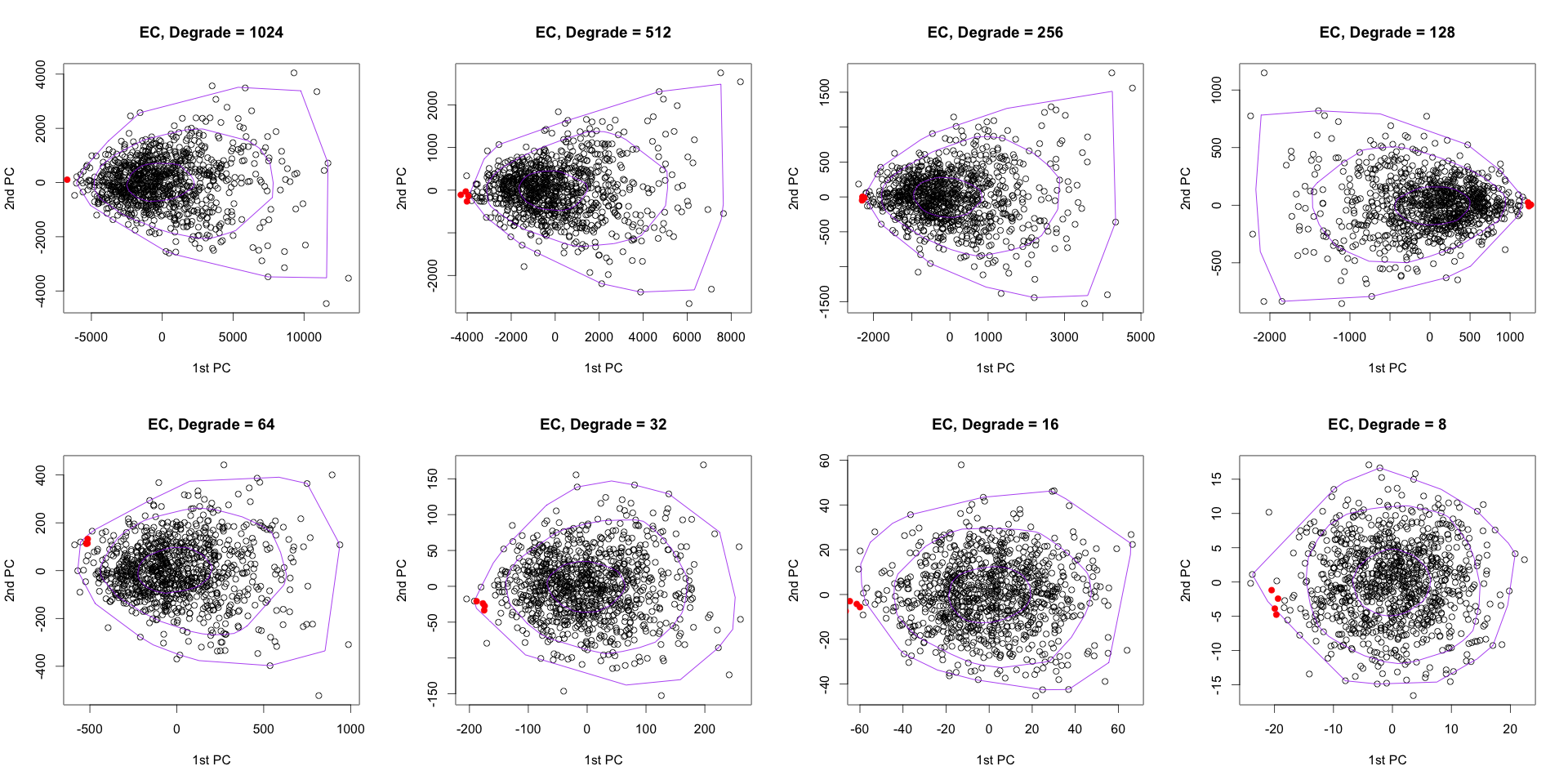}\\
	\caption{ Projection onto first two principal components for specific resolution tests for $\relEuler$.  Mahalanobis and depth contours corresponding to $p$-values of 0.1, 0.01 and 0.001 are shown in blue (top three rows) and purple (bottom three rows). Observed CMB points are in red.}
	\label{fig:specificContours_ec}
\end{figure*}

Regardless of the choice of test, it is evident that the model and observation disagree significantly at least in the number of loops on a range of scales approximately between 1 to 7 degrees.  For the Mahalanobis values, the general trend of significance increases up to N = 8, providing additional evidence the deviations are not purely due to chance. For both tests, the $p$-values for the summary tests tend to be more significant than the for individual resolutions. Another general trend is that the
non-parametric test shows the difference between the observed and the simulated maps to be starker than the parametric test. To help
interpret this difference, Figures~\ref{fig:globalContours}~--~\ref{fig:specificContours_ec} shows plots that visualize
to what extent the assumption of a Gaussian distribution for the
compared quantities is justified. See also Table~\ref{tab:degrade-pvalues_2} for a comparison with $p$-values for the absolute homology. The results indicate similar trends as in the relative homology case. 

\subsection{Principal component graphs}
\label{sec:global-pvalues}

Figure~\ref{fig:globalContours} presents the projection onto the first two principal components for the summary tests, which include results from all
resolutions. Mahalanobis and depth contours corresponding to $p$-values of 0.1, 0.01 and 0.001 are shown in blue (top) and purple (bottom). Observed CMB points are in red. Examining the diagrams corresponding to the Mahalanobis distance, the hypothesis that the distribution conforms to elliptical contours is questionable. 

Figures~\ref{fig:specificContours_b0}~--~\ref{fig:specificContours_ec}  present the projection onto the first two principal components respectively for $\relBetti{0}$, $\relBetti{1}$, and $\relEuler$ for specific resolutions. Also drawn are the Mahalanobis (top two rows) and Tukey depth (bottom two rows) contours. In general, it is the case that the symmetric Mahalanobis contours do not always fit the data. However, as the resolution decreases, the Mahalanobis contours, which are Gaussian in nature, seem to fit the data well, and may be a reasonable approximation after all.

\begin{table}
	\tiny \centering
	\begin{tabular}{|r||rrr|rrr|} \hline
		
		\multicolumn{7}{|c|}{Absolute homology} \\
		\hline
		
		& \multicolumn{3}{|c|}{$Mahalanobis$}  & \multicolumn{3}{|c|}{Tukey Depth} \\
		
		resolution & \multicolumn{1}{c}{$\Betti{0}$} & \multicolumn{1}{c}{$\Betti{1}$}
		
		& \multicolumn{1}{c}{$\Euler$}
		
		& \multicolumn{1}{|c}{$\Betti{0}$} & \multicolumn{1}{c}{$\Betti{1}$}
		
		& \multicolumn{1}{c|}{$\Euler$} \\ \hline \hline
		
		\multicolumn{7}{|c|}{threshold = 0.70} \\ \hline
		
		1024       & 0.416&	0.318	&0.629	&$\bf{<0.001}$&$\bf{<0.001}$&	0.501\\
		
		512       & 0.492	&0.371	&0.573		&0.241	&0.133	&0.437\\
		
		256       &  0.535	&0.308	&0.488		&0.131	&$\bf{<0.001}$&$\bf{<0.001}$\\
		
		128       &  0.490	&0.316	&0.542		&0.163	&$\bf{<0.001}$	&0.445\\
		
		64       &  0.413	&0.184	&0.396	&	0.180	&$\bf{<0.001}$&$\bf{<0.001}$\\
		
		32       &  \bf 0.012	&\bf 0.024	&\bf 0.035	& $\bf{<0.001}$&$\bf{<0.001}$&$\bf{<0.001}$\\
		
		16       & 0.220	&\bf0.017	&0.135		&0.164	&$\bf{<0.001}$&$\bf{<0.001}$\\
		
		8       &  $\bf{<0.001}$	&0.584	&\bf0.004		&$\bf{<0.001}$	&0.647	&$\bf{<0.001}$\\ \hline
		
		summary & 0.064	&0.276	&\bf 0.040		&0.152	&0.502	&$\bf{<0.001}$ \\ \hline\hline
		
		\multicolumn{7}{|c|}{threshold = 0.80} \\ \hline
		
		1024 	&  0.416	&0.318	&0.629&$\bf{<0.001}$	&$\bf{<0.001}$	&0.501	\\
		
		512   	&  0.492&	0.371	&0.573		&0.241	&0.133	&0.437\\
		
		256  	& 0.535	&0.308	&0.488		&0.131	&$\bf{<0.001}$&$\bf{<0.001}$\\
		
		128   	& 0.490	&0.316	&0.542		&0.163&$\bf{<0.001}$&	0.445\\
		
		64 	&  0.413	&0.184	&0.396	&	0.180&$\bf{<0.001}$	&$\bf{<0.001}$\\
		
		32 	&  \bf0.012	&\bf0.024	&\bf0.035&$\bf{<0.001}$&$\bf{<0.001}$&$\bf{<0.001}$ \\
		
		16    	&  0.220	&\bf0.017	&0.135		&0.164	&$\bf{<0.001}$	&$\bf{<0.001}$\\
		
		8    	&  \bf 0.001&	0.584	&\bf 0.004	&$\bf{<0.001}$&	0.647&$\bf	<0.001$\\ \hline
		
		summary	& $\bf{<0.001}$&	0.204&\bf	0.001	&$\bf{<0.001}$	&\bf0.010&\bf	0.001\\ \hline\hline
		
		\multicolumn{7}{|c|}{threshold = 0.90} \\ \hline
		
		1024 	& 0.281	& 0.339	& 0.642		& $\bf{<0.001}$	& $\bf{<0.001}$& 	0.301	\\
		
		512   	& 0.523	& 0.389	& 0.697	&	0.135	&$\bf{<0.001}$& 	0.520\\
		
		256  	& 0.402& 0.329& 	0.422	& $\bf{<0.001}$	& $\bf{<0.001}$	& 0.307\\
		
		128   	& 0.542	& 0.216	& 0.342	& 	0.153& $\bf{<0.001}$&	0.298\\
		
		64 	& 0.276& 	0.228& 	0.421	& $\bf{<0.001}$&$\bf{<0.001}$& 	0.448\\
		
		32 	&  \bf0.045& 	0.172	& 0.231		& $\bf{<0.001}$& 	0.175&$\bf{<0.001}$\\
		
		16    	& 0.058& 	\bf0.019& \bf	0.037& $\bf{<0.001}$& $\bf{<0.001}$& $\bf{<0.001}$	\\
		
		8    	& 0.223	& 0.475	& 0.071	& 	0.199	& 0.510	& $\bf{<0.001}$\\ \hline
		
		summary	&  \bf0.009& 	0.386& \bf	0.024		& 0.371	&0.631& 	0.206\\ \hline\hline

		\multicolumn{7}{|c|}{threshold = 0.95} \\ \hline
		
		1024 	& 0.281& 	0.339	& 0.642	& $\bf{<0.001}$& $\bf{<0.001}$& 	0.301	\\
		
		512   	& 0.523& 	0.407& 	0.698		& 0.250& 	$\bf{<0.001}$& 	0.512 \\
		
		256  	& 0.380& 	0.339& 	0.302	& 	0.132	& $\bf{<0.001}$& $\bf{<0.001}$\\
		
		128   	& 0.456& 	0.226	& 0.248		& $\bf{<0.001}$	& $\bf{<0.001}$& $\bf{<0.001}$\\
		
		64 	& 0.491	& 0.449	& 0.794		& 0.235	& 0.346	& 0.431	\\
		
		32 	&0.261& 	0.262	& 0.553	& $\bf{<0.001}$	& 0.308	& $\bf{<0.001}$	\\
		
		16    	& 0.137	& 0.066	& \bf0.050	& 	0.155&$\bf{<0.001}$& $\bf{<0.001}$\\
		
		8    	& 0.155	&\bf0.024	&\bf0.008	&$\bf{<0.001}$&$\bf{<0.001}$&\bf	$\bf{<0.001}$\\ \hline
		
		summary	& \bf0.049	& \bf0.033	& \bf0.036		&$\bf{<0.001}$	&$\bf{<0.001}$& $\bf{<0.001}$	\\ \hline

	\end{tabular}
	\caption{Table displaying the two-tailed $p$-values for absolute homology obtained from parametric (Mahalanobis distance) and non-parametric (Tukey depth) tests, for four mask binarization thresholds. The last entry for each threshold is the summary statistic computed across all resolutions. Marked in boldface are $p$-values $0.05$ or smaller.} 
	\label{tab:degrade-pvalues_2}
\end{table}

\begin{figure}
	\centering  
	\subfloat{\includegraphics[width=0.25\textwidth]{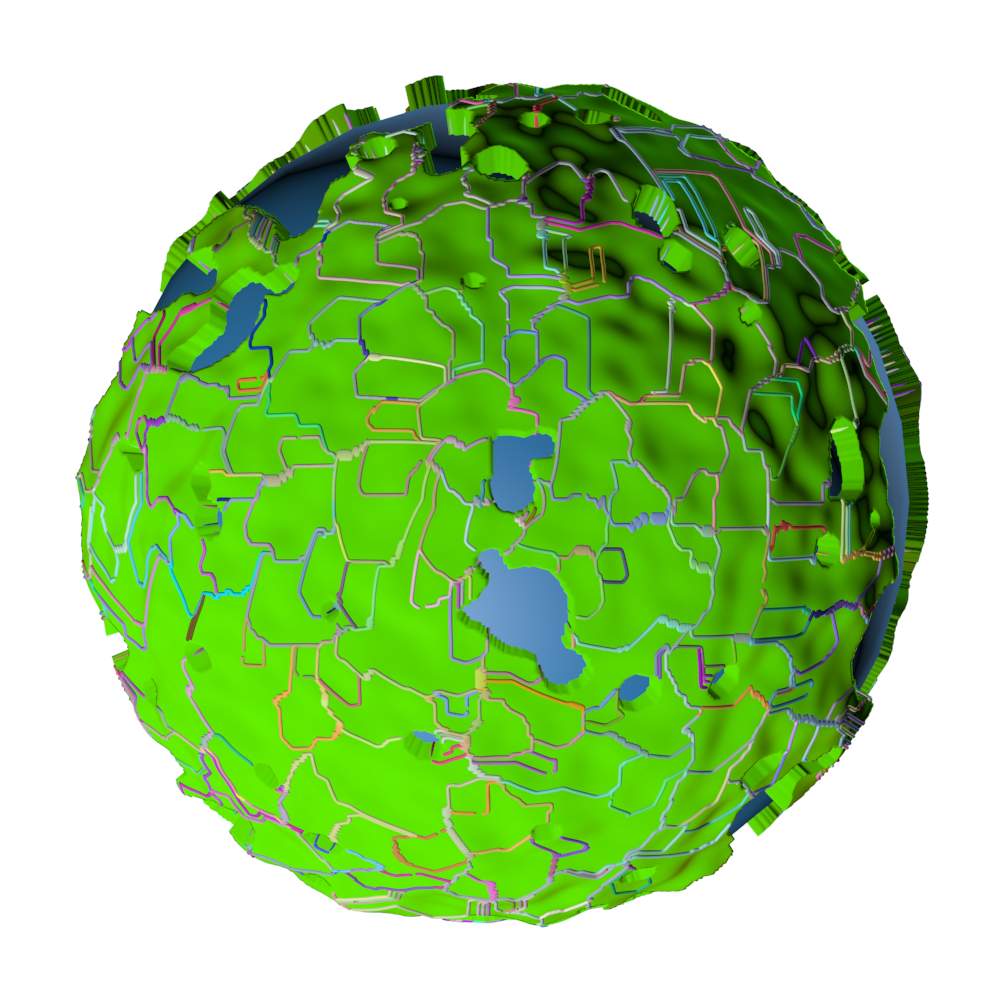}}
	\subfloat{\includegraphics[width=0.25\textwidth]{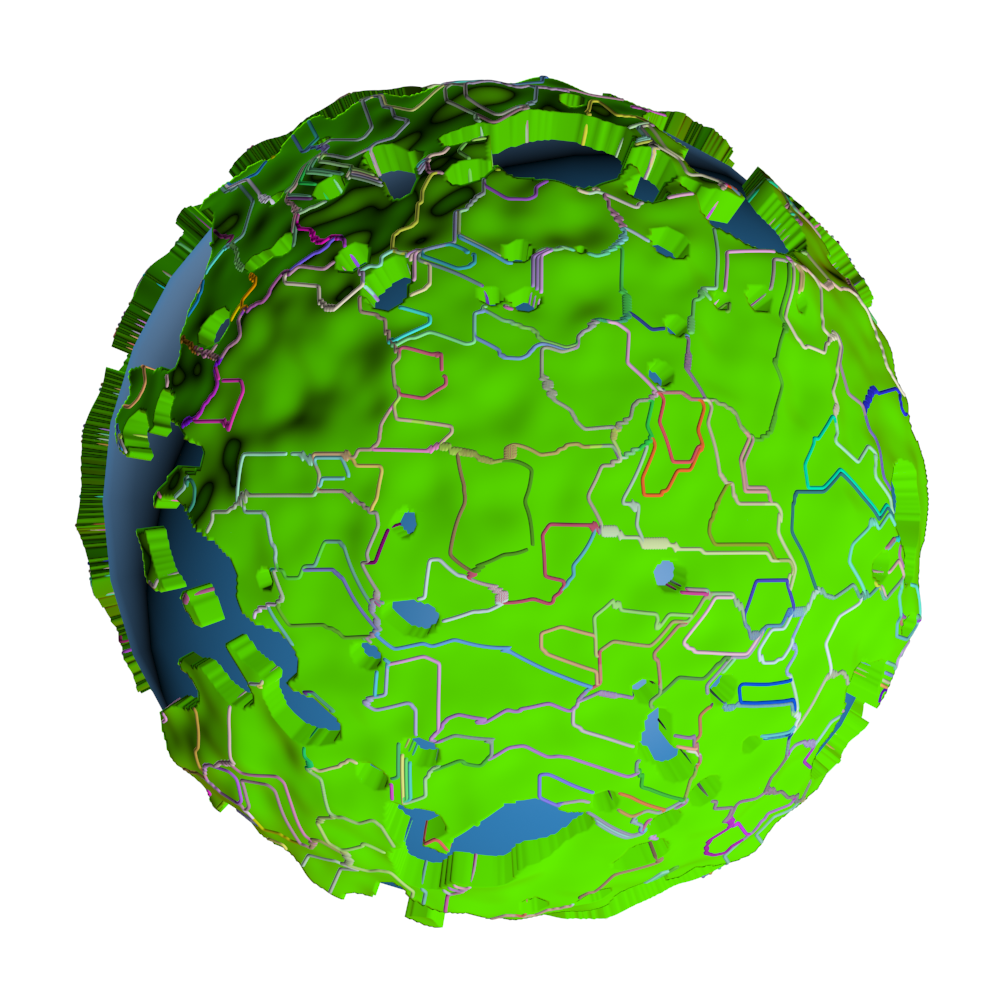}}\\
	\subfloat{\includegraphics[width=0.25\textwidth]{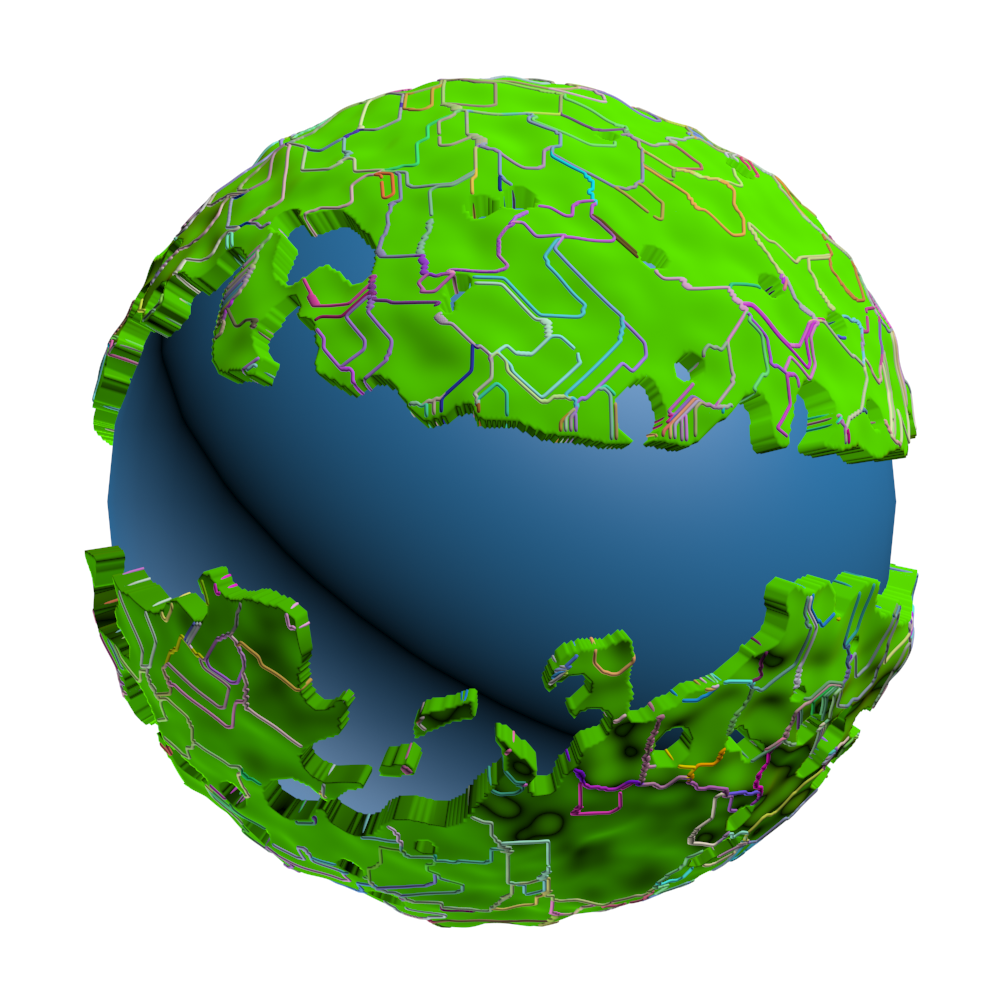}}
	\subfloat{\includegraphics[width=0.25\textwidth]{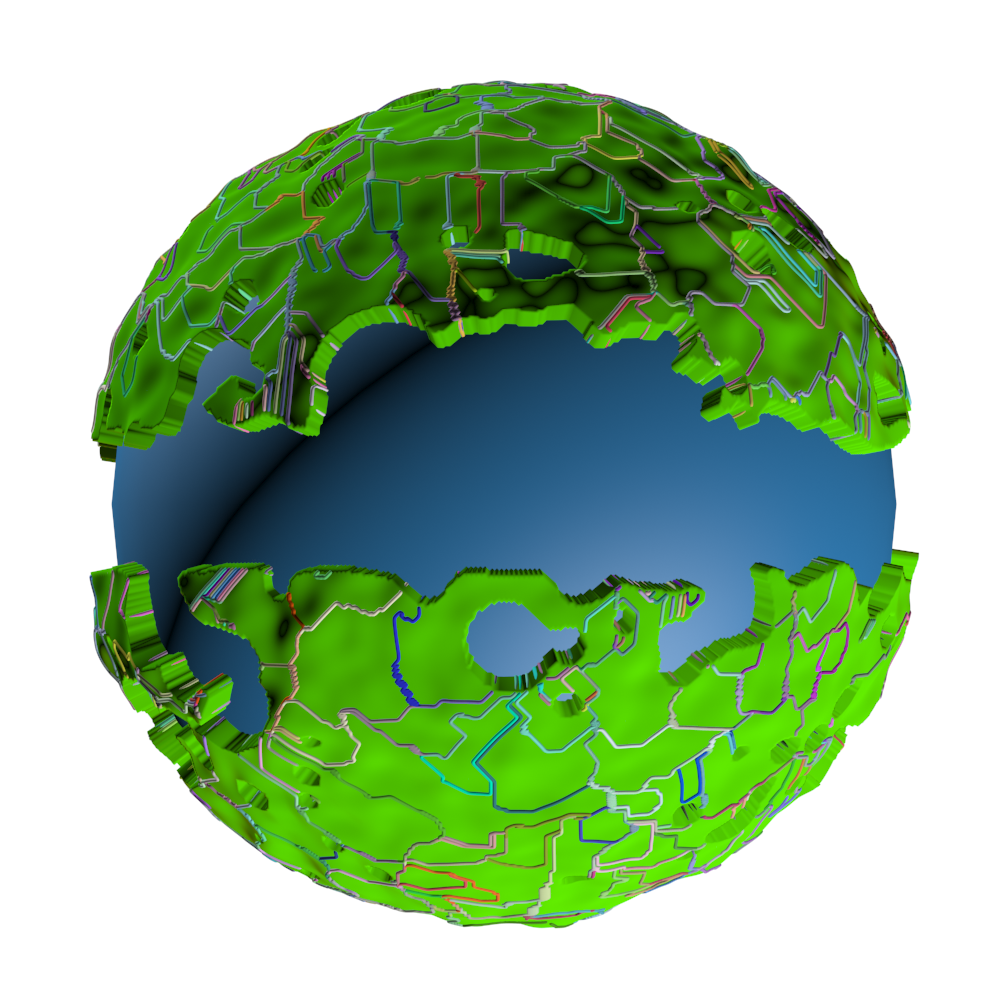}}\\
	\caption{Visualization of the loops for the largest excursion
		set, which consists of the entire sphere minus the mask. To improve
		the visualization, the temperature field has been smoothed by a small
		amount, and we do not draw very short loops.  From \emph{left} to \emph{right}: the sphere from the \emph{top},
		the \emph{bottom}, the \emph{left}, and the \emph{right} views.}
	\label{fig:geom_stats}
\end{figure}

\section{Summary and Conclusions}
\label{sec:discussion}

We provide evidence for the deviation of the observed Planck CMB maps from the Gaussian predictions of the standard LCDM model. Specifically, we find an over-abundance of loops in the observed maps,  deviating from the simulations at per cent to less than per mil levels. This is in terms of $p$-values computed using  $\chi^2$ statistics, between the resolutions $N = 32$ and $N = 8$.  The difference in the number of components and loops peaks sporadically at more than
$3\sigma$ from the predictions between $N = 32$ and $N = 8$. Results based on smoothed maps corroborate with those based on degraded maps in terms of approximate scales at which the anomaly is observed. We also compute the absolute homology for the dataset, and confirm that the results
are consistent with those from relative homology. External evidence that these deviations are not a result of overanalysing the data comes from the fact that the variance of the observed CMB is anomalous with respect to the standard model at $N = 16$ \citep{PlanckXXIII}, and the computed power spectrum exhibits a dip roughly at this range of scales. In addition, there are reports of a mildly significant Euler characteristic at $3.66$ degrees ($N = 16$) \citep{Eriksen04NG}, computed from independent measurements of the CMB by Planck's predecessor -- Wilkinson Microwave Anisotropy Probe (WMAP) satellite. This can be explained by the significantly high number of loops and components, together with cancellation effects that the Euler characteristic suffers from, by definition. Similar observations by independent satellites makes it unlikely that the source of the anomaly has its origin in instrumental noise or systematic effects. Moreover the medium super-horizon scales at we observe it, could possibly point to a cosmological origin. The non-parametric Tukey depth test shows the observations to be different from the simulations at almost all resolutions. Regardless of the preferred test,  evidently the topological structure of the CMB is deviant from the simulations, at least on some scale.

We can rule out this anomaly being the effect of the cold spot in the CMB sky, or any previously detected directional anomalies. Our statistics are based on a large numbers of loops surrounding the low density regions, to which the loop generated by the cold spot may contribute at most only a few, and often only one. Moreover, to support this claim, we visually confirm that these loops are scattered all over the sky (see Figure~\ref{fig:geom_stats}). We also test and
confirm that simulations that are based on Gaussian prescriptions and match the characteristics of the observed ``dipped'' power spectrum cannot resolve this anomaly. 
Additionally, we present topological methods that are suitable in the presence of obfuscating masks.  As such, the results presented in this paper are robust despite lacking full sky coverage, as well as being model-independent. 

In conclusion, we reiterate that we present clear evidence of departure of the observed CMB maps with respect to the simulations based on the LCDM paradigm, but make no attempt to address the issue of the physical mechanism behind this phenomenon; a question we leave to the wider cosmological community. Nevertheless, our analysis demonstrates the existence of unexpected topology in the CMB. Possible, but non-exhaustive scenarios worth exploring may be primordial non-Gaussianity, as well as models with non-trivial topology.

\section*{Acknowledgements}
PP and RA acknowledge the support of ERC advanced grant Understanding Random Systems through Algebraic Topology (URSAT) (no: 320422, PI: RA). This work is also part of a project that has received funding for PP and TB from the European Research Council (ERC)
under the European Union's Horizon 2020 research and innovation programme (grant agreement ERC advanced grant 740021--  Advances in Research on THeories of the dark UniverSe (ARTHUS), PI: TB).
HE and HW acknowledge the support by the Office of Naval Research, through grant N62909-18-1-2038, and by the DFG Collaborative Research Center TRR 109, `Discretization in Geometry and Dynamics', through grant I02979-N35 of the Austrian Science Fund (FWF).
PP is grateful to Julian Borill from the Planck consortium for providing the data, and for the illuminating discussions and inputs. PP also thanks Hans Kristen Eriksen and Anne Ducout for numerous discussions in the early stages.

\clearpage

\bibliographystyle{aa}
\bibliography{references.bib}

\end{document}